\documentclass[eqsecnum,aps,prd,nofootinbib]{revtex4}
\usepackage[dvips]{graphicx}

\usepackage{amsmath}
\usepackage{enumerate}
\usepackage{mathrsfs}
\usepackage{amsfonts}

\def\ben{\begin{equation}}
\def\een{\end{equation}}
\def\bena{\begin{eqnarray}}
\def\eena{\end{eqnarray}}

\newcommand{\I}{{\mathcal I}}

\newcommand{\F}{{\mathcal F}}

\newcommand{\tg}{{\tilde g}}
\newcommand{\tnabla}{{\tilde \nabla}}
\newcommand{\tn}{{\tilde n}}
\newcommand{\tgam}{{\tilde \gamma}}
\newcommand{\thh}{{\tilde h}}
\newcommand{\tP}{{\tilde P}}

\begin{document}
\title {Supersymmetric Multi-trace Boundary Conditions in AdS}

\author{Aaron J. Amsel\footnote{\tt amsel@physics.ucsb.edu} and Donald Marolf\footnote{\tt marolf@physics.ucsb.edu}}

\affiliation{Physics Department, UCSB, Santa Barbara, CA 93106, USA}

\begin{abstract}
Boundary conditions for massive fermions are investigated in
AdS${}_d$ for $d \ge 2$. For fermion masses in the range $0 \le |m|
< 1/2\ell$ with $\ell$ the AdS length, the standard notion of
normalizeability allows a choice of boundary conditions. As in the
case of scalars at or slightly above the Breitenlohner-Freedman (BF)
bound, such boundary conditions correspond to multi-trace
deformations of any CFT dual.  By constructing appropriate boundary
superfields, for $d=3,4,5$ we identify joint scalar/fermion boundary
conditions which preserve either ${\cal N}=1$ supersymmetry or
${\cal N}=1$ superconformal symmetry on the boundary.   In
particular, we identify boundary conditions corresponding via AdS/CFT (at large
$N$) to a 595-parameter family of double-trace marginal deformations of
the low-energy theory of $N$ M2-branes which preserve ${\cal N} =1$
superconformal symmetry.  We also establish that (at large $N$ and
large 't Hooft coupling $\lambda$) there are no marginal or relevant
multi-trace deformations of 3+1 ${\cal N} =4$ super Yang-Mills which
preserve even ${\cal N}=1$ supersymmetry.
\end{abstract}

\maketitle

\tableofcontents

\section{Introduction}

A central aspect of the Anti-de Sitter/Conformal Field Theory
(AdS/CFT) correspondence \cite{M, gkp, witten} is that
gauge-invariant deformations of the CFT Lagrangian correspond to
modifications of the AdS boundary conditions.  The requirement of
gauge invariance can be implemented by constructing operators that
transform in the adjoint representation of the gauge group and
taking a trace.  In this way, these deformations can be classified by
the number of traces.   Single-trace deformations, which in the CFT
correspond to the addition of simple sources, correspond
\cite{witten,kw} to fixing one of the Fefferman-Graham type coefficients \cite{FG} that control the fall-off of bulk fields at infinity.  In contrast, multi-trace deformations
correspond to imposing relations between two or more such coefficients
\cite{witten1}.

Of course, one should choose boundary conditions that lead to a
well-defined bulk theory.  In particular, they should be compatible
with the bulk inner product, ensuring that it is both finite and
conserved.  As observed in \cite{BF,kw}, for tachyonic scalars near
the Breitenlohner-Freedman bound, the standard scalar inner product
is in fact compatible with a variety of boundary conditions; see
\cite{BF,iw,mr,cm} for comments on the vector and tensor cases. It
is also possible to consider more general inner products \cite{cm}
(and see comments in \cite{mr}), though at the risk of sacrificing
positivity and introducing ghosts. We shall therefore restrict
attention to the standard inner products below\footnote{Strictly
speaking, in the context of non-linear theories one should speak of the symplectic
structure instead of the inner product. Since the symplectic
structure is simply an (indefinite) inner product on the space of
linearized fields, we will take this to be implied by our use of the
term ``inner product'' without further comment.}.

Our goal here is to understand supersymmetric multi-trace boundary
conditions for bulk scalar supermultiplets, generalizing certain
results of \cite{Marolf2006} for massless multiplets in AdS${}_4$.
Supersymmetric single-trace boundary conditions in AdS${}_4$ were analyzed in \cite{BF, hawking,in} (see also \cite{st} for the corresponding analysis in AdS${}_2$).
Some AdS/CFT applications of multi-trace boundary conditions were given in \cite{witten1, bss,ss}.  Supersymmetric  boundary conditions at finite boundaries were studied in \cite{vN}.

We begin our study by investigating multi-trace boundary conditions
for Dirac fermions of general mass $m$ propagating on a fixed
AdS${}_d$ background with $d \ge 2$.  See \cite{BF} for some
discussion of fermion boundary conditions in $d = 4$ and e.g.,
\cite{hs,frolov,henneaux99} for treatments of single-trace boundary
conditions for fermions in the context of AdS/CFT.  Our fermions are
free aside from possible non-linear boundary conditions.  We find
unique boundary conditions for $|m| \ge \frac{1}{2}$, but a wide
class of boundary conditions for $|m| < \frac{1}{2}$. One should be
able to include bulk interactions using the techniques of either
\cite{hmtz} or \cite{Amsel2006,Marolf2006}, and based on the results
of those works for scalars, one would not expect this to change the
allowed boundary conditions\footnote{However, in special cases
interactions do give rise to new logarithms which may break a
conformal invariance that appears to be preserved in the linearized
approximation.}.

We then specialize to the cases $d=3,4,5$ to study supersymmetry. We
study allowed boundary conditions for systems of scalars and
spin-1/2 fermions as specified in Table~\ref{tab1}.  For
$d=4,5$ these systems admit ${\cal N}=1$ bulk supersymmetry, while
our $d=3$ system admits ${\cal N}=(1,0)$ supersymmetry\footnote{In
fact, for $d=3$ our field content is that of an ${\cal N}=(2,0)$
supermultiplet, though we will only find non-trivial boundary
conditions which preserve ${\cal N}=(1,0)$. }. As in
\cite{Marolf2006}, constructing appropriate boundary superfields
from the Fefferman-Graham coefficients of bulk fields will allow us
to identify boundary conditions preserving either the full
supersymmetry (i.e., superconformal symmetry on the boundary) or a
certain subalgebra (naturally called boundary Poincar\'e
supersymmetry) containing half of the original supercharges.  With
our field content, non-trivial boundary conditions preserving
boundary Poincar\'e supersymmetry are generally allowed for fermion mass $|m| <
\frac{1}{2}$, though choices preserving superconformal invariance
(and which correspond to an integer number of traces in a dual field
theory) arise only for special values of $m$.  As in
\cite{Marolf2006}, we find that supersymmetric boundary conditions
always relate two distinct bulk scalar fields.  As a result, the
multi-trace boundary conditions allowed for a single scalar (and
used in so-called designer gravity theories \cite{thgh}) admit no
supersymmetric generalization.

\begin{table}

\begin{tabular}{|c|c| c| c|}
\hline
\, $d$ \, & \, Fermion Field Content \, &  \, \# of Real Scalars \, & \, \# of Real Supercharges \, \\
\hline
3\, & 2 Majorana & 2 & 2 \\
4\, & 1 Majorana & 2 & 4 \\
5\, & 1 Dirac & 4 &  8 \\
\hline
\end{tabular}
\vspace{.5cm}
\caption{ \label{tab1} Scalar supermultiplet content.  Note that in $d=5$, a complex Dirac spinor can be equivalently represented by a pair of symplectic Majorana spinors.}
\vspace{.2cm}
\end{table}

The plan of this paper is as follows.  After stating our conventions
in section~\ref{conventions}, we describe the allowed boundary
conditions in section \ref{BCs}.  Here we relegate technical details
to the appendices: appendix \ref{revion2} reviews solutions of the
AdS${}_d$ Dirac equation following \cite{ion2}, and appendix
\ref{normpsi} analyzes the convergence of the inner product.  We
then classify boundary conditions preserving supersymmetry as stated
above for AdS${}_4$ (section \ref{d4}), AdS${}_5$ (section
\ref{d5}), and AdS${}_3$ (section \ref{d3}). We close with some
discussion in section \ref{disc}, including comments on
supersymmetric deformations of M2-brane theories
\cite{BaggerLambert,ABJM} and 3+1 ${\cal N} =4$ super Yang-Mills.


\subsection{Conventions}
\label{conventions}

It is convenient to discuss boundary conditions using the conformal
compactification of AdS spacetime.  One may describe this
compactification by starting with the global AdS metric
\begin{equation}
\label{metricd} ds^2 = -(1+r^2) dt^2 +\frac{dr^2}{1+r^2} +r^2
d\Omega^2_{d-2} \,.
\end{equation}
Here $d\Omega^2_{d-2}$ is the line element of the unit sphere
$S^{d-2}$ and we have set the AdS length $\ell$ to one.  Introducing
the coordinate $\Omega$ via $r = \Omega^{-1} -\Omega/4$, one defines
an unphysical metric $\tilde g_{ab} = \Omega^2 g_{ab}$ which
satisfies
\begin{equation}
\label{unphysical} \widetilde{ds^2} = d\Omega^2 -
\left(1+\frac{1}{4} \Omega^2 \right)^2 dt^2+ \left(1-\frac{1}{4}
\Omega^2 \right)^2 d\Omega^2_{d-2} \,.
\end{equation}
The unphysical spacetime is thus a manifold with boundary $\I \cong
\mathbb{R} \times S^{d-2}$ at $\Omega = 0$.

In this spacetime, $\tilde n_a = \tilde \nabla_a \Omega$ coincides
with the unit normal to the boundary, where $\tilde \nabla_a$ is the
torsion-free covariant derivative compatible with $\tilde g_{ab}$.
It is also useful to define the orthogonal projector $\thh_{ab} =
\tg_{ab} - \tn_a \tn_b$, which at $\Omega =0$ becomes the induced
metric on the boundary
\begin{equation}
\thh_{ab} dx^a dx^b \,|_\I = -dt^2+ d\Omega^2_{d-2} \,;
\end{equation}
i.e., the Einstein static universe.  Indices on all tensor fields
with a tilde are raised and lowered with the unphysical metric
$\tg_{ab}$ and its inverse $\tg^{ab}$.

Our conventions for treating spinors are as follows.  Spacetime
indices are denoted by $a, b, \ldots$, while indices on a flat
internal space are denoted by $\hat a, \hat b, \ldots = \hat 0 ,
\hat 1, \hat 2, \ldots$.  In $d$ spacetime dimensions, the Dirac
spinor representation is $2^{[d/2]}$ dimensional, where $[x]$ is the
integer part of $x$. The flat-space gamma matrices are $2^{[d/2]}
\times 2^{[d/2]}$ matrices satisfying
\begin{equation}
\{\gamma_{\hat a}, \gamma_{\hat b}\} = 2 \eta_{\hat a \hat b} \,,
\end{equation}
where $\eta_{\hat a \hat b}$ is the metric of Minkowski space with
signature $(-++ \ldots)$. We also note that $(\gamma^{\hat
0})^\dagger = -\gamma^{\hat 0}$ and $(\gamma^{\hat k})^\dagger =
\gamma^{\hat k}$, with $\hat k = \hat 1, \hat 2, \ldots$. For a
given spacetime metric $g_{ab}$, we can define an orthonormal frame
$\{e^{\hat a}{}_{a} \}$ which satisfies $e^{\hat a}{}_a e^{\hat
b}{}_b \eta_{\hat a \hat b} = g_{ab}$. The curved space gamma
matrices are then given by $\gamma_a = e^{\hat a}{}_{a} \gamma_{\hat
a}$ and satisfy $\gamma_{(a}\gamma_{b)} = g_{ab}$.  Below, we take
our covariant derivative $\nabla_a$ to act on spinors as
\begin{equation}
\label{wdef} \nabla_a \psi = \partial_a \psi + \Gamma_a \psi \ \
{\rm where} \ \ \Gamma_a = \frac{1}{4} \, \omega_a{}^{\hat a \hat c}
\gamma_{[\hat a} \gamma_{\hat c]} \ \ {\rm and} \ \ -d e^{\hat a} =
\omega^{\hat a}{}_{\hat c} \wedge e^{\hat c} \,.
\end{equation}
Here  $\Gamma_a$ is the spin connection and $\omega_a{}^{\hat a \hat
c}$ are the rotation coefficients. We assume that all spinors are
anticommuting, and define the Dirac conjugate of a spinor $\psi$ to
be $\overline{\psi} = \psi^\dagger \gamma^{\hat 0}$.   Tildes will
denote quantities defined analogously in terms of $\tilde g_{ab}$;
e.g., $\tilde \gamma_{(a} \tilde \gamma_{b)} = \tilde g_{ab}$.


\section{Boundary conditions for scalars and fermions}
\label{BCs}

Let us consider free theories of scalars and Dirac fermions
propagating on a fixed AdS${}_d$ background.   For a set of
scalars $\phi_I$ of masses $m_{\phi_I}$ and Dirac fermions
$\psi_{\hat I}$ of masses $m_{\psi_{\hat I}}$, the Lagrangian is
 \begin{equation}
 \label{phipsi}
L = \sum_I \left( - \frac{1}{2} \nabla^a \phi_I \nabla_a \phi_I -
\frac{1}{2} m^2_{\phi_I} \phi_I^2 \right) + \sum_{\hat I}
 i\left[ \frac{1}{2} \left(\overline \psi_{\hat I} \gamma^a
\nabla_a \psi_{\hat I} - \overline{\nabla_a \psi}_{\hat I} \gamma^a
\psi_{\hat I} \right) - m_{\psi_{\hat I}} \overline\psi_{\hat I}
\psi_{\hat I} \right].
 \end{equation}
We wish to identify boundary conditions for which the standard inner
product is both finite and conserved on the space of linearized
solutions. These conditions suffice to yield a well-defined phase
space. In particular, they imply that charges corresponding to the
AdS isometries are well-defined and conserved. They also ensure that
the linearized quantum theory evolves unitarily.

Since we consider only linear fields and use the standard inner
product, we may identify normalizeable modes separately for each
field.  We first briefly recall the results for scalar fields using
the Klein-Gordon inner product.  Denoting the Breitenlohner-Freedman
bound by $m^2_{BF} = - \frac{(d-1)^2}{4}$, one finds two cases.  For
$m_{\phi_I}^2 \ge m^2_{BF} +1 $, there is a unique complete set of
normalizeable modes,  and any other (non-normalizeable) modes must
be fixed by the boundary conditions.

In contrast, much more general boundary conditions are allowed for
$m_{\phi_I}^2 < m^2_{BF} +1$, though the case $m_{\phi_I}^2 < m_{BF}^2$ is
usually ignored due to instabilities. The boundary conditions are
most simply expressed in terms of a Fefferman-Graham-type expansion
\cite{FG} of $\phi_I$. For $m_{\phi_I}^2 \neq m^2_{BF}$ we
have\footnote{Including gravitational backreaction would modify this
expansion \cite{hmtz, Amsel2006} if $4 \lambda_- \leq (d-1)$, which
can only be satisfied for $d \leq 4$. To include backreaction while
avoiding this regime, one would have to restrict the range of
$m_{\phi_I}$, though this is most likely only a technical
complication \cite{hmtz, Amsel2006}. In any case, we ignore such
backreaction here and consider propagation on a fixed spacetime.  }
\begin{equation}
\label{phi} \phi_I = \alpha_I \Omega^{\lambda_{I,-}} + \beta_I
\Omega^{\lambda_{I,+}}  + \dots, \ \ {\rm where} \ \ \lambda_{I,\pm}
= \frac{d-1 \pm \sqrt{(d-1)^2+ 4 m^2_{\phi_I} }}{2} \, .
\end{equation}
Here $\alpha_I, \beta_I$ are independent of $\Omega$, but can depend
on time and angles on the sphere. For $m_{\phi_I}^2 = m^{2}_{BF}$, the
roots~(\ref{phi}) are degenerate and the solution becomes
\begin{equation}
\label{phibf} \phi_I = \alpha_I \Omega^{\lambda}\,\textrm{log}\,\Omega
+ \beta_I \Omega^{\lambda} + \dots \ \ {\rm where} \ \ \lambda =
(d-1)/2.
\end{equation}
We refer to $\alpha_I, \beta_I$ as the boundary fields corresponding
to the bulk field $\phi_I$.

Normalizeability places no restriction on the boundary fields, but
we must still impose conservation. Considering two linearized
solutions $\delta_{1,2} \phi_I$ and the corresponding $\delta_{1,2}
\alpha_I, \delta_{1,2} \beta_I$,  a short calculation shows that the
flux through the boundary is
\begin{equation}
\label{wphi2} {\F}^\phi =  \sum_I \int_{\cal I} (\lambda_{I,+}-\lambda_{I,-})
\left(\delta_1 \beta_I \delta_2\alpha_I
- \delta_2 \beta_I \delta_1\alpha_I \right)  d^{d-1}S \,,
\end{equation}
where $d^{\,d-1}S$ is the integration element on $\I$.
This flux vanishes precisely when the boundary conditions restrict
$\alpha_I,\beta_I$ to a ``Lagrange submanifold'' in the space of
possible $(\alpha_I, \beta_I)$.  Such boundary conditions can often be
(locally) specified by choosing a function\footnote{We assume throughout this work that $W$ does not involve derivatives of fields along the boundary (see \cite{mr2, cm} for subtleties that arise when $W$ involves time derivatives).} $W(\alpha_I,x)$ and
requiring
\begin{equation}
\label{onebc}(\lambda_{I,+}-\lambda_{I,-})   \beta_I(x) = \frac{\partial W}{\partial \alpha_I}\,,
\end{equation}
where here $x \in \I$.
For theories with a dual CFT, this boundary condition corresponds to
adding a multi-trace term $W(\mathcal{O}_I, x)$ to the field theory
Lagrangian, where $\mathcal{O}_I$ is the operator dual to $\phi_I$ for
$W = 0$ boundary conditions.  Certain exceptional cases that will arise in later sections can be constructed as limits of (\ref{onebc}), but for our purposes may be better described by choosing a $W$ that depends on both $\alpha$'s and $\beta$'s.  As an example, with $I = 1,2$ we might choose $W = W(\alpha_1, \beta_2)$ and take
\begin{equation}
(\lambda_{2,+}-\lambda_{2,-}) \alpha_2 = -  \frac{\partial W}{\partial \beta_2} \, , \quad (\lambda_{1,+}-\lambda_{1,-}) \beta_1 = \frac{\partial W}{\partial \alpha_1} \,.
\end{equation}  Such boundary conditions again correspond to adding $W$ to the dual  CFT Lagrangian of the $W = 0$ theory.

The corresponding analysis for Dirac fermions (with the standard
inner product) is performed in appendix \ref{normpsi}. We make heavy use of \cite{ion2}, which solves the massive Dirac
equation in AdS${}_d$ (see appendix \ref{revion2} for a review). For $|m_{\psi_{\hat I}}| \ge 1/2$, there is a
unique complete set of normalizeable modes.  Any other
non-normalizeable modes must be fixed by the boundary condition.  In
contrast, much more general boundary conditions are allowed for
$|m_{\psi_{\hat I}}| < 1/2$.

For this latter case it is again convenient to introduce boundary
fields. In appendix \ref{normpsi}, we derive the asymptotic expansion
\begin{equation}
\label{solexpand} \psi_{\hat I} = \alpha^\psi_{\hat I} \,
\Omega^{\frac{d-1}{2}-m_{\psi_{\hat I}}}+\beta^\psi_{\hat I} \,
\Omega^{\frac{d-1}{2}+m_{\psi_{\hat I}}}+ \alpha'{}^\psi_{\hat I} \,
\Omega^{\frac{d+1}{2}-m_{\psi_{\hat I}}} + \beta'{}^\psi_{\hat I} \,
\Omega^{\frac{d+1}{2}+m_{\psi_{\hat I}}} + O(\Omega^{\frac{d+3}{2}-|m_{\psi_{\hat I}}|}) \, .
\end{equation}
Here $\alpha^\psi_{\hat I}, \beta^\psi_{\hat I}$ are again boundary
fields depending only on time and angles on the $S^{d-2}$. For later
convenience we have included certain sub-leading terms whose
coefficients $\alpha'{}^\psi_{\hat I}, \beta'{}^\psi_{\hat I}$ are
determined by $\alpha^\psi_{\hat I}, \beta^\psi_{\hat I}$. The
coefficients satisfy
\begin{eqnarray}
\label{projecta} \tP_+ \alpha^\psi_{\hat I} = 0 \, , \quad \tP_-
\alpha^\psi_{\hat I} = \alpha^\psi_{\hat I}, \quad \alpha'{}^\psi_{\hat I} &=& - \frac{1}{1- 2m_{\psi_{\hat I}}} \, \thh^{ab} \tgam_a \tnabla_b \alpha^\psi_{\hat I}, \\
\cr \label{projectb} \tP_- \beta^\psi_{\hat I} = 0 \, , \quad
\tP_+ \beta^\psi_{\hat I} = \beta^\psi_{\hat I}, \quad
 \beta'{}^\psi_{\hat I} &=&  \frac{1}{1+ 2m_{\psi_{\hat I}}} \,
\thh^{ab} \tgam_a \tnabla_b \beta^\psi_{\hat I} \, ,
\end{eqnarray}
where we have defined the radial projectors $\tP_\pm = \frac{1}{2}
\left(1 \pm \tn_a \tgam^a \right)$.  Note that when $m_{\psi_{\hat I}} < 0$,  the $\beta^\psi_{\hat I}$ term in (\ref{solexpand}) is actually the leading term in the asymptotic expansion.

It remains to impose conservation.  Inserting the asymptotic
expansion (\ref{solexpand}) into the fermion inner product (see
appendix \ref{normpsi}) and using (\ref{projecta}),
(\ref{projectb}), we find the fermionic contribution to the flux
through the boundary
\begin{equation}
\label{flux} \F^\psi = i \sum_{\hat I} \int_\I  \left[\left( \overline{\delta_1 \alpha^\psi_{\hat I}} \, \delta_2
\beta^\psi_{\hat I}  - \overline{\delta_1 \beta^\psi_{\hat I}} \, \delta_2 \alpha^\psi_{\hat I}  \right)
 -\Big(\delta_1 \leftrightarrow \delta_2 \Big) \right] d^{\,d-1}S \,.
 \end{equation}
Recall, however, that we are interested
in theories of the form (\ref{phipsi}) which contain both scalars
and fermions.  In the pure scalar case, it was not
necessary for the flux to vanish separately for each scalar field.
Instead, only the total flux was required to vanish. Similarly, when
both scalars and fermions are present, it is only the total flux
$\F \equiv \F^\phi + \F^\psi$ involving both types of fields that must vanish.
This occurs when the boundary conditions restrict the fields to what
one may call a ``Lagrange submanifold'' in the space of all
$(\alpha_I, \beta_I, \alpha^\psi_{\hat I}, \beta^\psi_{\hat I})$.  Again, this is often locally equivalent to choosing a
real function $W(\alpha_{I}, \alpha_{\hat I}^\psi,  \overline{\alpha_{\hat I}^\psi}, x)$ and
defining
\begin{equation}
\label{twobc}
(\lambda_{I, +} - \lambda_{I,-}) \beta_I(x) = \frac{\partial W}{\partial \alpha_I}\,, \ \ \
 -i \beta^\psi_{\hat I}(x) = \frac{\partial W}{\partial \overline{\alpha_{\hat I}^\psi}}\,.
\end{equation}
In other situations, one may wish to choose $W$ to depend on $\alpha$'s and $\beta$'s in analogy with
the scalar case.  In each instance, this corresponds to deforming the $W = 0$ dual CFT by adding $W$ to its action.
For simplicity, we now assume that all scalars satisfy $m_{\phi_I}^2
< m_{BF}^2 +1$ and all fermions satisfy $|m_{\psi_{\hat I}}| <
1/2$.

As a brief example, consider a theory with one Dirac fermion and no
scalars and suppose that we desire a linear boundary condition which
respects local Lorentz-invariance and translation-invariance on the
boundary. In even dimensions we must impose
 \begin{equation}
 \label{bceven}
 \beta^\psi = i q \gamma_{d+1} \alpha^\psi
\ \ {\rm where} \ \ \gamma_{d+1} = \frac{i^{\frac{d-2}{2}}}{d!}
\epsilon^{a_1 \ldots a_d} \gamma_{a_1} \ldots \gamma_{a_d}
 = i^{\frac{d-2}{2}} \gamma^{\hat 0} \gamma^{\hat 1} \ldots \gamma^{\hat{d-1}}
 \,,
 \end{equation}
  for some real $q$.
 Here $\gamma_{d+1}$ is the analogue of $\gamma_5$ in $d = 4$ and satisfies
 \begin{equation}
 \{\gamma_{d+1}, \gamma^a\} = 0 \,, \quad \gamma_{d+1}^\dagger = \gamma_{d+1} \,, \quad (\gamma_{d+1})^2 = 1 \,.
 \end{equation}
The Breitenlohner-Freedman boundary conditions for $d = 4$
\cite{BF} correspond to the particular choices $q = 0$ or $q =
\infty$. Since $ \gamma_{d+1} \tP_{\pm} = \tP_{\mp} \gamma_{d+1}$, the
boundary condition (\ref{bceven}) is consistent with
(\ref{projecta}), (\ref{projectb}).  In contrast,
 in odd dimensions, the matrix $\gamma_{d+1}$ is
proportional to the identity and commutes with $\tP_\pm$ so that
 (\ref{bceven}) implies $\alpha^\psi =\beta^\psi =0$.
Thus, in odd dimensions this theory has no non-trivial linear
boundary conditions with the desired properties. However, as shown
below, there are more possibilities with a greater number of
fermions.

As a final comment we mention that, at least for linearized fields,
the above analysis is equivalent to studying self-adjoint extensions
of the spatial wave operator. Such an approach was applied to
massive scalar fields and massless vector and tensor fields in
\cite{iw}.  The authors showed that a simple 2-parameter family of wave
operators sufficed to describe all of these fields (though there is
some subtlety associated with the choice of inner product in the tensor
case, see \cite{cm}). As is briefly mentioned in appendix
\ref{normpsi}, this approach can also be used for our fermions, and
the analysis again reduces to the wave operator studied in
\cite{iw}. Comparison with \cite{iw} explicitly shows that stability
issues of the sort that would arise for scalars with $m_\phi^2 <
m^2_{BF}$ cannot occur for Dirac fermions with real mass $m_\psi.$
Indeed, the relevant inequality is $(m_\psi \pm \frac{1}{2})^2 \ge
0$.  Thus, in some sense $m_\psi = \pm \frac{1}{2}$ is analogous to
saturating the Breitenlohner-Freedman bound, even though the
boundary conditions are unique for $|m_\psi| \ge \frac{1}{2}$.

\section{${\cal N}=1$ supersymmetry in $d = 4$}
\label{d4}

Consider the AdS${}_4$ theory of a Majorana fermion $\hat \psi$ and two
real scalars $(A,B)$ with Lagrangian (\ref{phipsi}).  To match the usual normalization of the action for Majorana fermions we define $\psi \equiv \sqrt{2} \, \hat \psi$ and work exclusively with this rescaled spinor. The fermion
$\psi$ obeys the Majorana condition $\overline{\psi} = \psi^T C$,
where $C$ is the charge conjugation matrix and satisfies
\begin{equation}
\label{charge4} C \gamma^a C^{-1} = - (\gamma^a)^T \,, \quad C^T =
C^{-1} = C^\dagger = -C \,.
\end{equation}
When the scalars masses ($m_A,m_B$) and fermion mass ($m$) are
related by
\begin{equation}
\label{ma} m^2_A = m^2+m - 2, \quad
 m^2_B = m^2-m - 2 \, ,
\end{equation}
the action is invariant \cite{BF,is} under the $\mathcal{N} = 1$
supersymmetry transformations
\begin{eqnarray}
&\delta_\eta A = \frac{i}{\sqrt{2}} \, \overline{\eta} \psi, \quad
\quad \quad
\delta_\eta B = -\frac{1}{\sqrt{2}} \, \overline{\eta} \gamma_5 \psi& \\
&\delta_\eta \psi = -\frac{1}{\sqrt{2}} \, \left[ \gamma^a \nabla_a
(A + i \gamma_5 B) + (m-1) A +i (m+1) \gamma_5 B \right] \eta& \, ,
\end{eqnarray}
where the supersymmetry-generating parameter $\eta$ is a Killing
spinor, i.e.
\begin{equation}
\left(\nabla_a +\frac{1}{2} \gamma_a \right) \eta = 0 \, .
\end{equation}
The case $m=0$ was studied in \cite{Marolf2006}; we closely follow
their analysis and correct certain equations below.

An important feature of supersymmetry in anti-de Sitter space (see
e.g. \cite{dewit}) is that fields in the same multiplet do not
necessarily have the same mass, though the degeneracy is restored in
the flat space limit $\ell^{-1} \to 0$.
\begin{figure}
\begin{picture}(0,0)
\put(-80,-10){$m^2_B$} \put(69,-10){$m^2_A$} \put(122, -42){$m$}
\put(116, -104){$m^2_{BF}+1$} \put(116, -154){$m^2_{BF}$}
\end{picture}
\begin{center}
\includegraphics[width=3.5in]{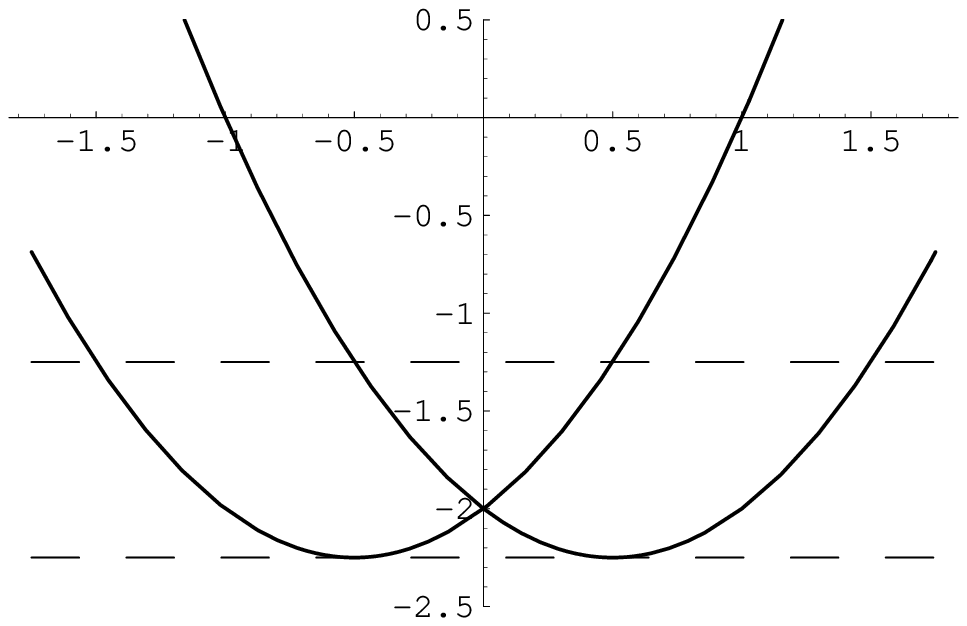}
\end{center}
\caption{The scalar masses $m^2_A, m^2_B$ (solid curves) are plotted
against the fermion mass $m$.  Dashed lines mark the window between
$m^2_{BF}$ and $m^2_{BF}+1$.} \label{massfig}
\end{figure}
The scalar masses are plotted in Fig.~\ref{massfig}. Notable
features of the relation (\ref{ma}) are as follows: {\it i\,}) massless
fermions correspond to conformally coupled scalar fields, $m_A^2 =
m_B^2 = -2$; {\it ii\,})  $m_{A,B}^2$ has a global minimum of $-9/4 =
m^2_{BF}$ at $m = \mp 1/2$, so the scalars always satisfy the
Breitenlohner-Freedman bound; {\it iii\,}) $m_{A,B}^2$ reaches the value
$m^2_{BF}+1$ at $m = \pm 1/2$.  Thus the range $|m| < 1/2$, which we
have seen allows general boundary conditions for fermions, typically
matches the mass range that allows general boundary conditions for
scalars.  The one exception occurs for $m = \pm 1/2$ where one
scalar saturates the BF bound and the other has squared mass
$m_{BF}^2 +1$.  The analogous properties also hold for the
AdS${}_5$, AdS${}_3$ cases studied in sections \ref{d5} and \ref{d3}.
We restrict attention to the case $|m| < 1/2$ below.

Solutions to the Killing spinor equation \cite{BF,Henneaux85} have
leading terms as $\Omega \to 0$ given by
\begin{equation}
\label{eta4} \eta = \eta_+ \Omega^{-1/2} +\frac{1}{2} \eta_- \Omega^{1/2}
+\ldots\,,
\end{equation}
where $\eta_\pm = \tP_\pm U$ and $U$ is a constant spinor. Using the
asymptotic expansions of $A,B, \psi, \eta$ from
\eqref{phi},\eqref{solexpand},\eqref{eta4} in the supersymmetry transformations and matching terms order
by order in $\Omega$ gives the action of supersymmetry on our
boundary fields:
\begin{eqnarray}
\label{bsusy4}
\delta_\eta \alpha_A &=& \frac{i}{\sqrt{2}} \, \overline{\eta_+} \, \alpha_\psi \\
\delta_\eta \beta_A &=& \frac{i}{2\sqrt{2}} \, \overline{\eta_-} \,
\beta_\psi +
\frac{1}{1+2m} \frac{i}{\sqrt{2}} \, \overline{\eta_+} \, \thh^{ab} \tgam_a \tnabla_b \beta_\psi \\
\delta_\eta \alpha_B &=& -\frac{1}{\sqrt{2}} \, \overline{\eta_+} \, \gamma_5 \beta_\psi \\
\delta_\eta \beta_B &=& -\frac{1}{2\sqrt{2}} \, \overline{\eta_-} \,
\gamma_5 \alpha_\psi + \frac{1}{1-2m} \frac{1}{\sqrt{2}} \,
\overline{\eta_+} \, \gamma_5 \thh^{ab} \tgam_a \tnabla_b
\alpha_\psi
\end{eqnarray}
\begin{eqnarray}
\delta_\eta \alpha_\psi &=& -\frac{1}{\sqrt{2}}\left[ (m-1) \alpha_A \eta_- - i(1- 2m) \gamma_5 \beta_B \eta_+ +\thh^{ab} \tgam_a \tnabla_b \alpha_A \eta_+ \right] \\
\label{bsusy4last} \delta_\eta \beta_\psi &=&
-\frac{1}{\sqrt{2}}\left[ i(m+1) \gamma_5 \alpha_B \eta_- + (1+ 2m)
\beta_A \eta_+ -i \gamma_5 \thh^{ab} \tgam_a \tnabla_b \alpha_B
\eta_+ \right] \,.
\end{eqnarray}

We now consider a 2-dimensional space of the 4 linearly independent
Killing spinors $\eta$ associated with some choice of Poincar\'{e}
coordinates for AdS:
\begin{equation}
\label{poincare} ds^2 = \frac{1}{z^2} \left(-dt^2+dz^2 +dx_1^2
+dx_2^2  \right)\,, \quad z\geq 0.
\end{equation}
In general, the conformally rescaled Killing spinor equation is
\begin{equation}
\label{conkilling} \left(\tnabla_a + \Omega^{-1} \tgam_a \tP_-
\right) \tilde \eta =0 \,,
\end{equation}
where $\tilde \eta = \Omega^{1/2} \eta$.  For the metric
(\ref{poincare}), we choose the conformal factor $\Omega = z$.
Solutions to (\ref{conkilling}) are then $\eta = \Omega^{-1/2}
\varepsilon_+$, where $\varepsilon_+$ is a constant spinor
satisfying $\tP_- \varepsilon_+ = 0$ and $\tP_\pm  =
\frac{1}{2}(1\pm \gamma_z)$, with $\gamma_z$ a flat space gamma
matrix.  For this two-dimensional space of Killing spinors, the
transformations (\ref{bsusy4})-(\ref{bsusy4last}) simplify somewhat,
and anti-commutators of such transformations generate the manifest
Poincar\'e symmetries of (\ref{poincare}).

To further simplify the supersymmetry transformations, note that the
4-component bulk Majorana spinors $\alpha_\psi, \beta_\psi,
\varepsilon_+$ satisfy projection conditions defined by $\tP_\pm$.
As a result, they define real, 2-component spinors living on the
boundary $\mathbb{R}^3$.  To make this explicit, let the indices
$i,j$ run over $t,x_1,x_2$.  Then the matrices $\underline{\Gamma}^j
= i \tgam^j \gamma_5$ satisfy $\{\underline{\Gamma}^i,
\underline{\Gamma}^j\} = \eta^{ij}$,
 $[\underline{\Gamma}^j , \tP_\pm] = 0$, and so form a representation of the 3-dimensional Clifford algebra on the boundary.  This is conveniently
 realized in terms of the real $2\times2$ matrices $\underline{\gamma}^t = i\sigma^2, \underline{\gamma}^{x_1} = \sigma^1, \underline{\gamma}^{x_2} = \sigma^3$, where
 $\vec{\sigma}$ denotes the Pauli matrices (\ref{pauli}).
 Our two-component spinor conventions are as follows.  Spinor indices are denoted by Greek letters $\kappa, \lambda, \ldots = 1,2$.  Spinor indices
 are raised and lowered with the antisymmetric tensors $\epsilon^{\kappa \lambda}, \epsilon_{\kappa \lambda}$, which we define by
 $\epsilon^{12} = 1 = -\epsilon^{21}, \epsilon_{\kappa \lambda} = -\epsilon^{\kappa \lambda}$.  Then
 \begin{equation}
 \psi^\kappa = \epsilon^{\kappa \lambda} \psi_\lambda \,, \quad  \psi_\kappa = \epsilon_{\kappa \lambda} \psi^\lambda
 \end{equation}
 and the spinor product is
 \begin{equation}
 \chi \psi \equiv \chi^\lambda \psi_\lambda = -\chi_\lambda \psi^\lambda = \psi^\lambda \chi_\lambda = \psi \chi \,.
 \end{equation}
In this standard notation, repeated spinor indices are summed over $1,2$.  Note that $(\underline{\gamma}^t)^{\kappa \lambda}  =
\epsilon^{\kappa \lambda}$, so for real spinors $\overline{\psi}
\chi = \psi^T \underline{\gamma}^t \chi = - \psi \chi$.  Also, the
three-dimensional Majorana condition $\bar \psi = \psi^T C$ reduces
exactly to the reality condition $\psi = \psi^*$, where the
three-dimensional charge conjugation matrix is $C =
\underline{\gamma^{t}}$.  Lastly, for Majorana spinors one has
\begin{equation}
\overline{\psi} \chi = \overline{\chi} \psi \,, \quad
\overline{\psi} \underline{\gamma}^j \chi = - \overline{\chi}
\underline{\gamma}^j \psi \,.
\end{equation}
Using these results, the Poincar\'e supersymmetries defined by $\eta
= \Omega^{1/2} \varepsilon_+$ may be written
\begin{eqnarray}
\label{t11}
\delta_\varepsilon \alpha_A &=& \frac{1}{\sqrt{2}} \,\varepsilon_+ \, a_\psi \\
(1+ 2m) \delta_\varepsilon \beta_A &=& -\frac{1}{\sqrt{2}} \,\varepsilon_+ \, \underline{\gamma}^j \partial_j \beta_\psi \\
\delta_\varepsilon \alpha_B &=& -\frac{1}{\sqrt{2}} \, \varepsilon_+ \,\beta_\psi \\
(1-2m) \delta_\varepsilon \beta_B &=& -\frac{1}{\sqrt{2}} \, \varepsilon_+ \, \underline{\gamma}^j \partial_j a_\psi \\
\delta_\varepsilon a_\psi &=& -\frac{1}{\sqrt{2}}\left[(1- 2m) \beta_B \varepsilon_+ - \underline{\gamma}^j \partial_j\alpha_A \varepsilon_+ \right] \\
\label{t22} \delta_\varepsilon \beta_\psi &=&
-\frac{1}{\sqrt{2}}\left[ (1+ 2m) \beta_A \varepsilon_+ +
\underline{\gamma}^j \partial_j \alpha_B \varepsilon_+ \right] \, ,
\end{eqnarray}
where $a_\psi$ denotes the two-component boundary spinor defined by
$i \gamma_5 \alpha_\psi$.

Under (\ref{t11})-(\ref{t22}), boundary fields mix only within each
of the disjoint sets ($\alpha_A, \alpha_\psi, \beta_B$), ($\alpha_B,
\beta_\psi, \beta_A$).  We may therefore construct useful boundary
superfields from each set separately.  To do so, introduce a real
anti-commuting 2-component spinor $\theta^\lambda$ and define
\begin{equation}
\Phi_- = \alpha_A + \overline{\theta} a_\psi +\frac{1}{2}
\overline{\theta} \theta (1-2m) \beta_B, \quad {\rm and} \quad
\Phi_+ = \alpha_B - \overline{\theta} \beta_\psi -\frac{1}{2}
\overline{\theta} \theta (1+2m) \beta_A \,.
\end{equation}
Taking $\theta$ to have conformal dimension $-1/2$, we note that the superfield $\Phi_\pm$
has a well-defined conformal dimension $1 \pm m$.
One may now check that (\ref{t11})-(\ref{t22}) can be written as
\begin{equation}
\label{sfxform4} \delta_\varepsilon \Phi_\pm = \frac{1}{\sqrt{2}}
\left[-\varepsilon_+^\kappa \frac{\partial}{\partial \theta^\kappa}
+\varepsilon_+ \underline \gamma^j \theta \partial_j \right]
\Phi_\pm \,
\end{equation}
and that $\delta_\varepsilon$ acts in precisely the same way on
$\delta_{\varepsilon} \Phi_\pm$; i.e., the $\Phi_\pm$ are indeed
superfields and (\ref{sfxform4}) defines a covariant derivative on
superspace.  Finally, using the above relations and the
two-component spinor identity $(\theta \psi) (\theta \chi)  =
-\frac{1}{2} (\theta \theta) \psi \chi$,  the total flux $\F^\phi + \F^\psi$ can
be written
\begin{equation}
\F = \int_\I d^{\, 3}S \int d^{\, 2}\theta \left(\delta_1 \Phi_-
\delta_2 \Phi_+ - \delta_1 \Phi_+ \delta_2 \Phi_-  \right) \, .
\end{equation}
It is now clear that for any function $W$, the boundary condition
\begin{equation}
\label{sfbc4} \Phi_- = \frac{\delta W(\Phi_+)}{\delta \Phi_+}
\end{equation}
conserves the inner product and is invariant under the Poincar\'e
supersymmetries. Such boundary conditions correspond to deformations
of a dual CFT action by the addition of a term $\int_\I d^{\, 3}S \int
d^{\, 2}\theta W.$

In terms of component fields we have
\begin{eqnarray}
\label{4comp}
\alpha_A &=& W'(\alpha_B) \\
(1-2m) \beta_B &=& -(1+2m) W''(\alpha_B) \beta_A +\frac{1}{2} W'''(\alpha_B) \overline{\beta_\psi} \gamma_5 \beta_\psi \\
\alpha_\psi &=& i W''(\alpha_B) \gamma_5 \beta_\psi \,.
\end{eqnarray}
For general $W$, these boundary conditions break the conformal (and
thus superconformal) symmetry; however, for the special choice
\begin{equation}
W(\Phi_+) = \frac{1+m}{2} \, q \left(\Phi_+ \right)^{\frac{2}{1+m}}
\end{equation}
the full $O(3,2)$ symmetry is preserved and, as one can explicitly
check, so is the full supersymmetry defined by
(\ref{bsusy4}-\ref{bsusy4last}) for the arbitrary Killing spinors
$\eta$.  I.e., superconformal symmetry is preserved on the boundary.

As a simple example, consider the linear boundary condition $\Phi_-
= q \Phi_+$.  In terms of the component fields this is
\begin{equation}
\label{lincompex} \alpha_A = q \alpha_B\,, \quad (1-2m) \beta_B = -q
(1+ 2m) \beta_A \,, \quad \alpha_\psi = i q \gamma_5 \beta_\psi \,.
\end{equation}
The boundary conditions of \cite{BF} correspond to $q =0$ or $q =
\infty$.  Note that (\ref{lincompex}) and, more generally,
(\ref{4comp}) relate the two scalar fields to each other ($\alpha_A$
to $\alpha_B$), ($\beta_A$ to $\beta_B$).  Boundary conditions of
the sort studied in e.g. \cite{witten1,thgh} which relate $\alpha_A$
to $\beta_A$ cannot be supersymmetrized to respect Poincar\'e
supersymmetry on the boundary.

\section{${\cal N}=1$ supersymmetry in $d = 5$}
\label{d5}

For $d=5$ we will consider a theory with four real scalars and a
single Dirac spinor written as a symplectic Majorana pair.  To set
up the notation for symplectic spinors, we choose the
five-dimensional gamma matrix representation $ \gamma^{\hat 0} =
\left(
\begin{matrix}
0 & -iI_2 \\
-iI_2 & 0
\end{matrix}
\right) , \quad \gamma^{\hat k} = \left(
\begin{matrix}
0 & -i\sigma^{k}
\\ i\sigma^{k} & 0
\end{matrix}\right), \quad \gamma^{\hat 4} = \left(
\begin{matrix}
-I_2& 0 \\
0 & I_2
\end{matrix}
\right)$  where $k= 1,2,3$.  We also define the charge conjugation
matrix $C = \left(
\begin{matrix}
i \sigma^2 && 0 \\
0 && i \sigma^2
\end{matrix}
\right) \,$, which satisfies $C^{-1} = C^\dagger = C^T = -C$ and $C
\gamma^a C^{-1} = +(\gamma^a )^T \,$. The sign difference between
this last relation and (\ref{charge4}) means that one cannot
consistently define Majorana spinors in $d=5$.  Instead, one can
impose a modified or ``symplectic'' Majorana condition (see e.g.
\cite{pvn})
\begin{equation}
\label{sm} \overline{\psi}^{\, i} = \psi_j^T \lambda^{ji} C \,,
\quad i,j = 1,2
\end{equation}
where the Dirac conjugate is
\begin{equation}
\overline{\psi}^{\,i} \equiv \psi^\dagger_i \gamma^{\hat 0}
\end{equation}
and $\lambda$ is the symplectic matrix
\begin{equation}
\lambda = \left(
\begin{matrix}
0 && 1 \\
-1 && 0
\end{matrix}
\right) \,, \quad \lambda^2 = -1, \quad \lambda^T = -\lambda.
\end{equation}
The symplectic indices are raised and lowered by $\lambda$, with the
convention
\begin{equation}
\psi^i = \psi_j \lambda^{ji} = -\psi_j \lambda^{ij} \,, \quad
\lambda^{ij} = \lambda_{ij} \,.
\end{equation}
With these definitions one can obtain the useful symplectic Majorana
flip formulas
\begin{equation}
\overline{\psi}^{\,i} \chi_j = -\lambda_{jk} \lambda_{il}
\overline{\chi}^{\,k} \psi_l \,, \quad \overline{\psi}^{\,i}
\gamma^a \chi_j = -\lambda_{jk} \lambda_{il} \overline{\chi}^{\,k}
\gamma^a \psi_l \,.
\end{equation}

Now, consider a Dirac spinor $\psi$ in $d = 5$. The fields $ \psi_1
\equiv \psi \,, \psi_2 \equiv -\gamma^{\hat 0} C \psi_1^*$
form a symplectic Majorana pair satisfying (\ref{sm}).  The
equations of motion are
\begin{equation}
\label{dirac5}
\gamma^a \nabla_a \psi_1 - m \psi_1 = 0 \,, \quad \gamma^a \nabla_a
\psi_2 + m \psi_2 = 0 \,, \quad {\rm or} \ \ \gamma^a \nabla_a \psi_i
-m M_{ij} \psi_j = 0 \,,
\end{equation}
where $M \in$ USp(2) is given by
\begin{equation}
M = \left(
\begin{matrix}
1 && 0 \\
0 && -1
\end{matrix}
\right) \,.
\end{equation}
In \cite{shuster}, the author considers theories with $M$ a general
element of USp(2), but for our purposes it will be sufficient to
consider the simplest case when the fermion mass matrix is diagonal.
This last equation of motion can also be obtained directly from the
Lagrangian
\begin{equation}
\label{sympL} L = -\frac{i}{2}\left(\overline{\psi}^{\, i} \gamma^a
\nabla_a \psi_i - m M_{ij} \overline{\psi}^{\,i} \psi_j \right) \, ,
\end{equation}
where the above reality condition relates $\psi_1$ and $\psi_2$.

In five dimensions, a (complex) Dirac spinor has four degrees of
freedom on-shell, and so via supersymmetry should correspond to four
real scalars.  Let us then consider a theory \cite{shuster} of our
standard form \eqref{phipsi} where $I = 1,\ldots 4$.  We take just
one Dirac fermion, which we think of as being described by the
symplectic Lagrangian \eqref{sympL} (note the extra overall minus sign relative to (\ref{phipsi}) inserted to match certain conventions of \cite{shuster}).  If the masses are related by
\begin{equation}
m^2_1 = m^2_2 = m^2+m -\frac{15}{4},  \ \quad m^2_3 = m^2_4 = m^2-m
-\frac{15}{4} \, ,
\end{equation}
the action is invariant under the ${\cal N}=1$ supersymmetry transformations
\begin{eqnarray}
\delta_\eta \phi^I &=& - i (\sigma^I \lambda)_{ij} \overline{\eta}^{\, i} \psi_j \\
\delta_\eta \psi_i &=& (\lambda \sigma^I)_{ji} \gamma^a \nabla_a
\phi^I \eta_j + \left(3/2 \, (\sigma^I)^T \lambda M +m M
\lambda (\sigma^I)^\dagger \right)_{ij} \, \eta_j \phi^I \,.
\end{eqnarray}
Here $\sigma^I = (\vec{\sigma}, i I_2)$ and $\eta_i$ is a symplectic
Majorana Killing spinor \cite{shuster} satisfying
\begin{equation}
\nabla_a \eta_i +\frac{1}{2} \gamma_a M_{ij} \eta_j = 0 \,.
\end{equation}
As in the four-dimensional case, massless fermions correspond to
conformally coupled scalar fields, $m_I^2 =-15/4$, and the scalars
always satisfy the BF bound, $m^2_I \geq m^2_{BF} = -4$. For $I =
1,2$ ($I = 3,4$), the BF bound is saturated at $m = -1/2$ ($m =
1/2$), and the scalar mass $m^2_{BF}+1$ is reached at $m = 1/2$ ($m
= -1/2$).  As usual, we consider only the range $-1/2 < m < 1/2$
below.

We again take the fermions $\psi_i$ to have the asymptotic form (\ref{solexpand}).  For $i=1$, the expansion coefficients satisfy the relations (\ref{projecta}), (\ref{projectb}). For $i=2$ however, these relations are slightly modified due to the opposite sign of the mass term in the Dirac equation (\ref{dirac5}).  In particular, we note that $\tP_- \alpha^\psi_2 = 0, \tP_+ \beta^\psi_2 = 0$ and that certain sign changes occur in the expressions for $\alpha'^\psi_2, \beta'^\psi_2$.  It is these properties  under the radial projectors that allow us to have non-trivial boundary conditions for
fermions in this odd dimensional theory, even though $\gamma_{d+1}$
is proportional to the identity.  We see that we can
consistently relate $\alpha^\psi_1$ to $\beta^\psi_2$ and
$\alpha^\psi_2$ to $\beta^\psi_1$, since they are the same ``type''
of spinor.
Solutions to the Majorana Killing spinor equation have leading terms
as $\Omega \to 0$ given by
\begin{equation}
\eta_i = \alpha^\eta_i \Omega^{-1/2} +\frac{1}{2} \beta^\eta_i
\Omega^{1/2} +\ldots
\end{equation}
where
\begin{equation}
\tP_- \alpha^\eta_1 = 0\,, \quad \tP_+ \beta^\eta_1 = 0, \quad \tP_+
\alpha^\eta_2 = 0\,, \quad \tP_- \beta^\eta_2 = 0 \, .
\end{equation}

 Inserting the asymptotic expansions of $\phi^I,
\psi_i, \eta_i$ into the supersymmetry transformations and matching
terms order by order in $\Omega$ we obtain the action of
supersymmetry on the boundary fields:
\begin{eqnarray}
\label{bsusy5}
\delta_\eta z_1 &=& i \overline{\alpha^\eta}^{\,1} \alpha^\psi_1 \\
\delta_\eta z_2 &=& -\frac{i}{2} \overline{\beta^\eta}^{\, 2} \alpha^\psi_1 +\frac{i}{1-2m}  \overline{\alpha^\eta}^{\,2} \thh^{ab} \tgam_a \tnabla_b \alpha^\psi_1 \\
\delta_\eta z_3 &=& -i \overline{\alpha^\eta}^{\,2} \beta^\psi_1 \\
\delta_\eta z_4 &=& \frac{i}{2} \overline{\beta^\eta}^{\,1} \beta^\psi_1 +\frac{i}{1+2m}  \overline{\alpha^\eta}^{\,1} \thh^{ab} \tgam_a \tnabla_b \beta^\psi_1 \\
\delta_\eta z_1^\dagger &=& -i \overline{\alpha^\eta}^{\,2} \alpha^\psi_2 \\
\delta_\eta z_2^\dagger &=& -\frac{i}{2} \overline{\beta^\eta}^{\,
1} \alpha^\psi_2
-\frac{i}{1-2m}  \overline{\alpha^\eta}^{\,1} \thh^{ab} \tgam_a \tnabla_b \alpha^\psi_2 \\
\delta_\eta z_3^\dagger &=& -i \overline{\alpha^\eta}^{\,1} \beta^\psi_2 \\
\delta_\eta z_4^\dagger &=& -\frac{i}{2} \overline{\beta^\eta}^{\,2}
\beta^\psi_2
+\frac{i}{1+2m}  \overline{\alpha^\eta}^{\,2} \thh^{ab} \tgam_a \tnabla_b \beta^\psi_2 \\
\delta_\eta \alpha^\psi_1 &=& 2\left(m-\frac{3}{2}\right) z_1 \beta^\eta_1 +2(1-2m)z_2 \alpha^\eta_2 +2 \thh^{ab} \tgam_a \tnabla_b z_1 \alpha^\eta_1 \\
\delta_\eta \alpha^\psi_2 &=& 2\left(m-\frac{3}{2}\right)
z_1^\dagger \beta^\eta_2 -2(1-2m)z_2^\dagger \alpha^\eta_1
-2 \thh^{ab} \tgam_a \tnabla_b z_1^\dagger \alpha^\eta_2 \\
\delta_\eta \beta^\psi_1 &=& -2\left(m+\frac{3}{2}\right) z_3 \beta^\eta_2 +2(1+2m)z_4 \alpha^\eta_1 -2 \thh^{ab} \tgam_a \tnabla_b z_3 \alpha^\eta_2 \\
\label{bsusy5last} \delta_\eta \beta^\psi_2 &=&
2\left(m+\frac{3}{2}\right) z_3^\dagger \beta^\eta_1
+2(1+2m)z_4^\dagger \alpha^\eta_2
 -2 \thh^{ab} \tgam_a \tnabla_b z_3^\dagger \alpha^\eta_1 \,,
\end{eqnarray}
where we have defined the complex boundary scalars
\begin{eqnarray}
z_1 &=& \frac{1}{2} (\alpha_1 + i\alpha_2) \,, \quad z^\dagger_1 = \frac{1}{2} (\alpha_1 - i\alpha_2) \\
z_2 &=& \frac{1}{2} (\beta_3 + i\beta_4) \,, \quad z^\dagger_2 = \frac{1}{2} (\beta_3 - i\beta_4) \\
z_3 &=& \frac{1}{2} (\alpha_3 + i\alpha_4) \,, \quad z^\dagger_3 = \frac{1}{2} (\alpha_3 - i\alpha_4) \\
z_4 &=& \frac{1}{2} (\beta_1 + i\beta_2) \,, \quad z^\dagger_4 =
\frac{1}{2} (\beta_1 - i\beta_2) \,.
\end{eqnarray}

We now consider a subspace of the set of Killing spinors $\eta$
associated with some choice of Poincar\'{e} coordinates
\begin{equation}
ds^2 = \frac{1}{z^2} \left(-dt^2+dz^2 +dx_1^2
+dx_2^2+dx_3^2  \right)\,, \quad z\geq 0 \,.
\end{equation}
Solutions to the conformal Killing spinor equation are then $\eta_i = \Omega^{-1/2}
\varepsilon_i$, where $\varepsilon_i$ are constant spinors
satisfying $\tP_- \varepsilon_1 = 0, \tP_+ \varepsilon_2 = 0$.

The 4-component bulk spinors $\alpha^\psi_i, \beta^\psi_i, \varepsilon_i$ satisfy projection conditions defined by $\tP_\pm$.  As a result, they define 2-component spinors
living on the boundary $\mathbb{R}^4$.  To make this explicit, let the index $\bar j$ run over $t, \vec{x}$ and take the
four-dimensional Dirac matrices to be
\begin{equation}
\gamma^{\bar j} = (\gamma^{\hat 0}, \gamma^{\hat k }) = \left(
\begin{matrix}
0 && -i \sigma^{\bar j} \\
-i \bar \sigma^{\bar j} && 0
\end{matrix}
\right) \,,
\end{equation}
where we have defined $\sigma^{\bar j} = (I_2, \vec{\sigma})$ and
$\bar \sigma^{\bar j} =(I_2, -\vec{\sigma})$.  Now, the radial
projectors are $\tP_\pm = \frac{1}{2} (1 \pm \gamma_z)$ and it is
natural to choose $\gamma_z = \gamma^{\hat 4}$. Note however, that
$\gamma^{\hat 4} = i \gamma^{\hat 0} \gamma^{\hat 1} \gamma^{\hat 2}
\gamma^{\hat 3}$ which serves as the ``boundary $\gamma_5$.''  So,
the radial projectors match onto chiral projectors on the boundary,
and then a four-component bulk spinor in five dimensions that has been acted on
with $\tP_\pm$ gets mapped to a two-component left or right-handed
Weyl spinor in four dimensions. In particular, we have
\begin{eqnarray}
\alpha_1^\psi = \left(
\begin{matrix}
\alpha_\kappa \\
0
\end{matrix}
\right) \,, \quad \alpha_2^\psi = \left(
\begin{matrix}
0 \\
i \alpha^{\dagger \dot \kappa}
\end{matrix}
\right) \,, \quad
\beta_1^\psi = \left(
\begin{matrix}
0 \\
\beta^{\dagger \dot \kappa}
\end{matrix}
\right) \,, \quad \beta_2^\psi = \left(
\begin{matrix}
-i \beta_\kappa \\
0
\end{matrix}
\right)
\end{eqnarray}
\begin{eqnarray}
\varepsilon_1 = \left(
\begin{matrix}
0 \\
\varepsilon^{\dagger \dot \kappa}
\end{matrix}
\right) \,, \quad \varepsilon_2 = \left(
\begin{matrix}
-i \varepsilon_\kappa \\
0
\end{matrix}
\right) \,.
\end{eqnarray}
Our conventions for two-component spinors are the same as in the
previous section.  We have further introduced conjugate spinors and
dotted spinor indices, $(\chi_\kappa)^\dagger = \chi^\dagger_{\dot
\kappa}$.  Dotted indices are raised and lowered by contracting with
the second index of $\epsilon^{\dot \kappa \dot \lambda},
\epsilon_{\dot \kappa \dot \lambda}$, with $\epsilon^{\dot 1 \dot 2}
=1 = -\epsilon_{\dot 1 \dot 2}$.  We also note the index placement
$\sigma^{\bar j} =  \sigma^{\bar j}_{\kappa \dot \lambda}$ and $\bar
\sigma^{\bar j} =\bar \sigma^{\bar j \dot \kappa \lambda}$.  The
calculations below use the identity $\psi^\dagger \bar \sigma^{\bar
j} \chi = - \chi \sigma^{\bar j} \psi^\dagger$.

In this notation, the Poincar\'{e} supersymmetries defined by $\eta_i = \Omega^{-1/2}
\varepsilon_i$ may be written
\begin{eqnarray}
\label{susy5}
\delta_\varepsilon z_1 &=& \varepsilon \alpha \\
(1-2m)\delta_\varepsilon z_2 &=& \varepsilon^\dagger \bar \sigma^{\bar j} \partial_{\, \bar j} \alpha \\
\delta_\varepsilon \alpha &=& -2i(1-2m) z_2 \varepsilon -2i \sigma^{\bar j}  \partial_{\, \bar j} z_1 \varepsilon^\dagger \\
\delta_\varepsilon z_3^\dagger &=& i\varepsilon \beta \\
(1+2m)\delta_\varepsilon z_4^\dagger &=& -i \varepsilon^\dagger \bar \sigma^{\bar j} \partial_{\, \bar j} \beta \\
\label{susy5last} \delta_\varepsilon \beta &=& 2(1+2m) z_4^\dagger
\varepsilon -2 \sigma^{\bar j}  \partial_{\, \bar j} z_3^\dagger
\varepsilon^\dagger \,.
\end{eqnarray}
The transformations of the complex conjugate fields may be obtained
by taking the complex conjugate of the above relations.

Under (\ref{susy5})-(\ref{susy5last}), boundary fields mix only within each of the disjoint sets
$(z_1, z_2, \alpha)$, $(z_3^\dagger,
z_4^\dagger, \beta)$.  We may therefore construct useful boundary superfields from each set separately,
\begin{equation}
\Phi_1 = z_1 +\theta \alpha -i (1-2m) \theta \theta z_2 \ \quad
\Phi_2 = z_3^\dagger +i \theta \beta + i (1+2m) \theta \theta
z_4^\dagger \,.
\end{equation}
Taking the conformal dimension of $\theta$ to be $-1/2$, we note that the superfield $\Phi_{1,2}$ has
conformal dimension $\frac{3}{2} \mp m$.
One may now check that (\ref{susy5}-\ref{susy5last})
can be written as a superspace covariant derivative acting on superfields,
\begin{equation}
\label{sfxform5} \delta_\varepsilon \Phi = \left(\varepsilon^\kappa
\frac{\partial}{\partial \theta^\kappa}- 2 i \theta \sigma^{\bar j}
\varepsilon^\dagger \partial_{\, \bar j} \right) \Phi \,.
\end{equation}
 Similarly one can define the conjugate superfields
\begin{eqnarray}
&\Phi_1^\dagger = z_1^\dagger +\theta^\dagger \alpha^\dagger +i
(1-2m) \theta^\dagger \theta^\dagger z_2^\dagger, \ \quad
\Phi_2^\dagger = z_3 -i \theta^\dagger \beta^\dagger -i (1+2m)
\theta^\dagger \theta^\dagger z_4,& \cr
 &{\rm and } \quad \delta_\varepsilon
\Phi^\dagger = \left(\varepsilon^\dagger_{\dot \kappa}
\frac{\partial}{\partial \theta^\dagger_{\dot \kappa}}-
 2 i \theta^\dagger \bar \sigma^{\bar j} \varepsilon \partial_{\bar j} \right) \Phi^\dagger \, . &
\end{eqnarray}
Finally, using the above relations the total flux can be expressed as
\begin{equation}
\F = \left[-i \int_\I d^{\,4}S \int d^{\,2} \theta \left(\delta_1
\Phi_1 \delta_2 \Phi_2 -\delta_1 \Phi_2 \delta_2 \Phi_1 \right)
\right] + \Big[ \quad \Big]^\dagger \, .
\end{equation}
It is now clear that for any function $W$, the boundary condition
\begin{equation}
\label{sfbc5} \Phi_1 = i \, \frac{\delta W(\Phi_2)}{\delta \Phi_2}
\end{equation}
conserves the inner product and is invariant under the Poincar\'{e} supersymmetries.

In terms of the original component fields we have
\begin{equation}
z_1 = i W'(z_3^\dagger), \quad (1-2m) z_2= -i (1+2m)
W''(z_3^\dagger) z_4^\dagger +\frac{1}{4} W'''(z^\dagger_3)
\overline{\beta^\psi}^{\, 1} \beta^\psi_2, \quad \alpha^\psi_1 = - i
W''(z_3^\dagger) \beta^\psi_2 \,.
\end{equation}
For general $W$ these boundary conditions break the
conformal (and thus superconformal) symmetry; however, for the special choice
\begin{equation}
W(\Phi_2) = \frac{3/2+m}{3} \, q \left(\Phi_2 \right)^{\frac{3}{3/2+m}}
\end{equation}
the full $O(4,2)$ symmetry is preserved and so is
the full supersymmetry defined by (\ref{bsusy5}-\ref{bsusy5last})
for the arbitrary Killing spinors $\eta_i$.  I.e., superconformal symmetry is preserved on the boundary.

As a simple example, consider the
linear boundary condition $\Phi_1 = i q \Phi_2$.  In terms of the
component boundary fields this is
\begin{equation}
\alpha_1 = q \alpha_4 \,, \quad \alpha_2 = q \alpha_3 \,, \quad
\beta_4 = -q\, \frac{1+2m}{1-2m} \beta_1\,, \quad \beta_3 = -q\,
\frac{1+2m}{1-2m} \beta_2
\end{equation}
\begin{equation}
\alpha^\psi_1 = -i q \beta^\psi_2 \,, \quad \alpha^\psi_2 = -i q
\beta^\psi_1 \,.
\end{equation}
Note that these boundary conditions relate the two scalar fields to each other ($\alpha_{1,2}$ to $\alpha_{3,4}$),  ($\beta_{1,2}$ to $ \beta_{3,4}$).  As in the AdS${}_4$ theory treated above, boundary conditions
relating $\alpha_I$ to $\beta_I$ cannot be supersymmetrized to respect the  Poincar\'{e} supersymmetry on the boundary.

\section{${\cal N} = (1,0)$ supersymmetry in $d = 3$}
\label{d3}

In $2+1$ dimensions, the AdS supergroup has the factored form $G_L
\times G_R$, and so $AdS_3$ supergravity theories can be labeled as
having $\mathcal{N} = (p,q)$ supersymmetry \cite{at}.  In such
theories, a Majorana (real) spinor has one degree of freedom
on-shell, and so should correspond via supersymmetry to one real
scalar.  As noted above however, in odd dimensions there does not
seem to be a Lorentz invariant way of imposing generalized boundary
conditions on a single fermion without introducing derivatives.  So let us begin with a  theory containing two copies of a scalar multiplet. This theory in fact has ${\cal N} = (2,0)$ supersymmetry, but we will
only find interesting boundary conditions that preserve a $(1,0)$
subalgebra.   The theory \cite{it,dkss,dksst} consists of two
scalars $\phi_i = (\phi_1, \phi_2)$ and two Majorana fermions $\hat \psi^A =
(\hat \psi^1, \hat \psi^2)$ with Lagrangian \eqref{phipsi}, where
$m_{\phi} \equiv m_{\phi_1} = m_{\phi_2}$ and $m \equiv m_{\psi_1} =
m_{\psi_2}.$  To match the usual normalization of the action for Majorana fermions we define $\psi^A \equiv \sqrt{2} \, \hat \psi^A$ and work exclusively with this rescaled spinor.  The fermions $\psi^A$ obey the Majorana condition $\overline{\psi} = \psi^T C$, where $C
= i \sigma^2$ is the charge conjugation matrix.  With the real gamma matrix representation $\gamma^{\hat a} = (i \sigma^2, \sigma^1, \sigma^3)$, the Majorana condition
amounts to the reality condition $\psi = \psi^*$.  Other conventions
for 2-component spinors are the same as given above.

For
 \begin{equation}
 m^2_\phi = m^2+m -\frac{3}{4} \, ,
 \end{equation}
the action is invariant under the $\mathcal{N} = (2,0)$
supersymmetry transformations
\begin{eqnarray}
\delta_\eta \phi_i &=& \frac{i}{2}  \overline{\eta} \Gamma_i \psi \\
\delta_\eta \psi &=& -\frac{i}{\sqrt{2}} \left[\gamma^a \nabla_a
\phi_i \Gamma_i \eta + \left(m-\frac{1}{2}\right)\phi_i \Gamma_i
\eta \right] \, .
\end{eqnarray}
Here we have suppressed the spinor labels $A,B, \ldots = 1,2$ and
the matrices $\Gamma^{AB}_i$ are $\Gamma_1 = \sigma^1, \Gamma_2 =
\sigma^3$. The supersymmetry-generating parameters $\eta^A$ are Majorana Killing spinors
satisfying
\begin{equation}
\nabla_a \eta^A +\frac{1}{2} \gamma_a \eta^A = 0 \,.
\end{equation}

As in the higher dimensional cases treated above, massless fermions
correspond to conformally coupled scalar fields, $m_\phi^2 =-3/4$,
and the scalars always satisfy the BF bound, $m^2_\phi \geq m^2_{BF}
= -1$.  The BF bound is saturated at $m = - 1/2$, and $m_\phi^2$ reaches $m^2_{BF}+1$ at $m = 1/2$.  We once again
restrict attention to the range $-1/2 < m < 1/2$.  The theory studied in
\cite{it} is a limiting case of the theory in \cite{dkss},
corresponding to choosing $m = 1/2$.

Solutions to the Killing spinor equation have leading
terms as $\Omega \to 0$ given by
\begin{equation}
\eta^A = \eta^A_+ \Omega^{-1/2} +\frac{1}{2} \eta^A_- \Omega^{1/2}
+\ldots\,,
\end{equation}
where $\eta^A_\pm = \tP_\pm U^A$ and $U^A$ is a constant spinor.
We expect that valid boundary conditions will relate $\alpha_\psi^1$
to  $\alpha_\psi^2$ and $\beta_\psi^1$ to $\beta_\psi^2$, since
these spinors exhibit the same properties under the radial
projectors.

Inserting the asymptotic expansions of $\phi_i, \psi^A, \eta^A$ into
the supersymmetry transformations and matching terms order by order
in $\Omega$ gives the action of supersymmetry on our boundary fields \cite{dksst}:

\begin{eqnarray}
\label{bsusy3}
\delta_\eta \alpha_i &=& \frac{i}{\sqrt{2}} \, \overline{\eta_+} \, \Gamma_i \alpha_\psi \\
\delta_\eta \beta_i &=& \frac{i}{2\sqrt{2}} \, \overline{\eta_-} \,
\Gamma_i \beta_\psi +
\frac{1}{1+2m} \frac{i}{\sqrt{2}} \, \overline{\eta_+} \, \Gamma_i \thh^{ab} \tgam_a \tnabla_b \beta_\psi \\
\delta_\eta \alpha_\psi &=& -\frac{1}{\sqrt{2}}\left[ \left(m-\frac{1}{2}\right) \alpha_i \Gamma_i \eta_-
+\thh^{ab} \tgam_a \tnabla_b \alpha_i \Gamma_i \eta_+ \right] \\
\delta_\eta \beta_\psi &=& -\frac{1}{\sqrt{2}} (1+ 2m) \beta_i
\Gamma_i \eta_+ \,.
\end{eqnarray}
Here we shall only attempt to preserve the $(1,0)$ supersymmetry
transformations\footnote{The full $(2,0)$ transformations can be preserved
with Dirichlet or Neumann type boundary conditions \cite{dksst}; whether this can be done with more general boundary conditions as well
is a matter for further investigation.  It is not immediately obvious how to do so, since the $(2,0)$ boundary supersymmetry multiplets are
$(\alpha_i, \alpha_\psi^A)$, $(\beta_i, \beta_\psi^A)$, and thus the natural boundary conditions for the spinors would lead to
relating fields in the same multiplet.} given by setting $\eta^1 = 0, \eta^2 \equiv \eta$.
We then have
\begin{eqnarray}
\label{susy3}
\delta_\eta \alpha_1 &=& \frac{i}{\sqrt{2}} \, \overline{\eta_+} \, \alpha_\psi^1 \\
\delta_\eta \alpha_2 &=& -\frac{i}{\sqrt{2}} \, \overline{\eta_+} \, \alpha_\psi^2 \\
\delta_\eta \beta_1 &=& \frac{i}{2\sqrt{2}} \, \overline{\eta_-}
\,\beta_\psi^1 +
\frac{1}{1+2m} \frac{i}{\sqrt{2}} \, \overline{\eta_+} \, \thh^{ab} \tgam_a \tnabla_b \beta_\psi^1 \\
\delta_\eta \beta_2 &=& -\frac{i}{2\sqrt{2}} \, \overline{\eta_-}
\,\beta_\psi^2 -
\frac{1}{1+2m} \frac{i}{\sqrt{2}} \, \overline{\eta_+} \, \thh^{ab} \tgam_a \tnabla_b \beta_\psi^2 \\
\delta_\eta \alpha_\psi^1 &=& -\frac{1}{\sqrt{2}}\left[ \left(m-\frac{1}{2}\right) \alpha_1 \eta_- +\thh^{ab} \tgam_a \tnabla_b \alpha_1 \eta_+ \right] \\
\delta_\eta \alpha_\psi^2 &=& \frac{1}{\sqrt{2}}\left[ \left(m-\frac{1}{2}\right) \alpha_2 \eta_- +\thh^{ab} \tgam_a \tnabla_b \alpha_2 \eta_+ \right] \\
\delta_\eta \beta_\psi^1 &=& -\frac{1}{\sqrt{2}} (1+ 2m) \beta_1 \eta_+ \\
\label{susy3last} \delta_\eta \beta_\psi^2 &=& \frac{1}{\sqrt{2}}
(1+ 2m) \beta_2 \eta_+ \,.
\end{eqnarray}
We now consider a subspace
of the set of Killing spinors $\eta$ associated with some choice of
Poincar\'{e} coordinates
\begin{equation}
ds^2 = \frac{1}{z^2} \left(-dt^2+dz^2 +dx_1^2 \right)\,, \quad z\geq 0 \,.
\end{equation}
Solutions to the Killing spinor
equation are then $\eta = \Omega^{-1/2} \varepsilon_+$, where
$\varepsilon_+$ is a constant spinor satisfying $\tP_- \varepsilon_+
= 0$.

Let the index $\mu$ run over $t, x_1$ and take the two-dimensional Dirac matrices to be
$\gamma^\mu = (\gamma^{\hat 0 }, \gamma^{\hat 1}) = (i\sigma^2,
\sigma^1)$. Now, the radial projectors are $\tP_\pm = \frac{1}{2} (1
\pm \gamma_z)$ and it is natural to choose $\gamma_z = \gamma^{\hat
2}$. Note however, that $\gamma^{\hat 2} = \gamma^{\hat 0}
\gamma^{\hat 1} = \sigma^3$, which serves as the ``boundary
$\gamma_5$.''  So, the radial projectors match onto chiral
projectors on the boundary, and then a three-dimensional bulk spinor
that has been acted on with $\tP_\pm$ gets mapped to a Weyl
spinor in two dimensions.  Note that for any two-component spinors $\chi_\pm
\equiv \tP_\pm \chi, \psi_\pm \equiv \tP_\pm \psi$, we have
$\chi_\pm \psi_\pm = 0$.

Under (\ref{susy3})-(\ref{susy3last}), boundary fields mix only within each of the disjoint sets $(\alpha_1, \alpha^1_\psi)$, $(\alpha_2, \alpha^2_\psi)$,
$(\beta_1, \beta^1_\psi)$, and $(\beta_2, \beta^2_\psi)$.
 This suggests that we define the scalar boundary superfields
\begin{equation}
\label{scSF} \Phi_1 = \alpha_1 + \overline{\theta_+} \alpha_\psi^1,
\quad \Phi_2 = \alpha_2 - \overline{\theta_+} \alpha_\psi^2
\end{equation}
and the spinor boundary superfields
\begin{equation}
\label{spSF} \Psi_1 = \beta^2_\psi -i (1+2m) \theta_+  \beta_2,
\quad \Psi_2 = \beta^1_\psi +i (1+2m) \theta_+  \beta_1 \,.
\end{equation}
We again take $\theta_+$ to have conformal dimension $-1/2$ so that the scalar superfields both have conformal dimension $\frac{1}{2} -m$, while the spinor superfields both have conformal dimension $1+m$.
One may now check that (\ref{susy3}-\ref{susy3last}) (with $\eta_- \to 0, \eta_+ \to
\varepsilon_+$) can be written as a superspace covariant derivative acting on superfields,
\begin{equation}
\label{sfxform} \delta_\varepsilon \Phi = \frac{1}{\sqrt{2}} \left(i
\overline{\varepsilon_+} \frac{\partial}{\partial
\overline{\theta_+}}+
 \overline{\varepsilon_+}  \gamma^\mu \theta_+ \partial_\mu \right)
 \Phi,
 \quad
 \delta_\varepsilon \Psi = \frac{1}{\sqrt{2}} \left(i \varepsilon_+ \frac{\partial}{\partial\theta_+}+
 \overline{\varepsilon_+}  \gamma^\mu \theta_+ \partial_\mu \right) \Psi \,.
\end{equation}
Using relations \eqref{scSF} and \eqref{spSF}, the total
flux can be expressed as
\begin{equation}
\F = \left[-i \int_\I d^{\,2}S \int d \theta_+ \left(\delta_1 \Phi_1 \delta_2
\Psi_2 -\delta_1 \Phi_2 \delta_2 \Psi_1 \right) \right] - \Big[
\delta_1 \leftrightarrow \delta_2 \Big] \,.
\end{equation}
It is now clear that for any function $f$, the boundary condition
\begin{eqnarray}
\label{sfbc3} \Phi_1 = f(\Phi_2)\,,\qquad  \Psi_1 &=& \frac{\delta f(\Phi_2)}{\delta \Phi_2} \,
\Psi_2
\end{eqnarray}
conserves the inner product and is invariant under the
Poincar\'{e} supersymmetries.  These boundary conditions may be summarized through the spinor potential $\widetilde W = f(\Phi_2) \Psi_2$, in terms of which the deformation of any dual CFT action is $-i \int d^{\,2}S \int d\theta_+ \, \widetilde W$.

In terms of the component fields we have
\begin{eqnarray}
\label{bcs3}
\alpha_1 &=& f(\alpha_2) \\
(1+2m) \beta_2 &=& -(1+2m) f'(\alpha_2) \beta_1 + i f''(\alpha_2) \overline{\beta_\psi^1} \alpha^2_\psi \\
\alpha_\psi^1 &=& - f'(\alpha_2) \alpha^2_\psi \\
\label{bcs3last}
\beta_\psi^2 &=& f'(\alpha_2) \beta^1_\psi \,.
\end{eqnarray}
For general $\widetilde W$, these boundary conditions break the conformal (and thus superconformal)
symmetry.  However, for the special choice of linear boundary
conditions $\Phi_1 =  q \Phi_2\,, \Psi_1 =  q \Psi_2\, $, that is,
\begin{equation}
\alpha_1 = q \alpha_2 \,, \quad \beta_2 = -q \beta_1 \,, \quad
\alpha^1_\psi = -q \alpha^2_\psi \,, \quad \beta^2_\psi = q
\beta^1_\psi \,,
\end{equation}
the full AdS symmetry is preserved and so is the full supersymmetry
defined by (\ref{susy3}-\ref{susy3last}) for the arbitrary Killing
spinors $\eta$.  For this case, $\widetilde W$ has conformal
dimension $3/2$ and provides a marginal deformation of the dual CFT.
The result that the linear boundary conditions preserve
superconformal symmetry for any $|m| < 1/2$ is associated with
the two scalars always having equal masses; in the $d =4,5$ cases
considered above, the scalars masses coincide only when $m = 0$.

Alternatively, one can define two scalar superfields
\begin{equation}
\Xi_1 = \alpha_1 + \overline{\theta_+} \alpha_\psi^1 + i
\overline{\theta_-} \beta^2_\psi + (1+2m) \overline{\theta_-}
\theta_+ \beta_2, \quad \Xi_{2} = \alpha_2 - \overline{\theta_+}
\alpha_\psi^2 + i \overline{\theta_-} \beta^1_\psi - (1+2m)
\overline{\theta_-} \theta_+ \beta_1
\end{equation}
of conformal dimension $\frac{1}{2} - m$ (where we take $\theta_-$ to have conformal dimension $-\frac{1}{2}-2m$) and obtain the supersymmetry transformations by acting with the same
superspace derivative given in (\ref{sfxform}).  In terms of these
superfields, the flux can be expressed as
\begin{equation}
\F = \int_\I d^{\,2}S \int d \theta_+ d\overline{\theta_-} \left( \delta_1
\Xi_2 \delta_2 \Xi_1 - \delta_1 \Xi_1 \delta_2 \Xi_2\right) \,.
\end{equation}
The boundary condition $\Xi_1 = W'(\Xi_2)$ conserves the inner product and, when written out in terms of the component fields, gives the same expressions (\ref{bcs3}-\ref{bcs3last}) for $f = W'$ or $\widetilde W = -i \int d\overline{\theta_-} \, W$.

If instead we had set $\eta^1 = \eta, \eta^2 = 0$, we would have
obtained a different set of $\mathcal{N} =(1,0)$ supersymmetry
transformations, which can be obtained from those given above by
replacing $\alpha_\psi^1 \to \alpha_\psi^2\,, \alpha_\psi^2 \to
-\alpha_\psi^1\,, \beta_\psi^1 \to \beta_\psi^2\,,
\beta_\psi^2 \to -\beta_\psi^1$.  Performing a similar analysis in this case
leads to the general boundary conditions
\begin{eqnarray}
&\alpha_1 = W'(\alpha_2), \quad
(1+2m) \beta_2 = -(1+2m) W''(\alpha_2) \beta_1 - i W'''(\alpha_2) \overline{\beta_\psi^2} \alpha^1_\psi,& \\
&\alpha_\psi^2 = W''(\alpha_2) \alpha^1_\psi, \quad \beta_\psi^1 =
-W''(\alpha_2) \beta^2_\psi \,.&
\end{eqnarray}

\section{Discussion}
\label{disc}

Our study began with a general analysis of boundary conditions,
consistent with finiteness and conservation of the standard inner
product, for Dirac fermions in AdS spacetime (as had been previously
done for bosonic fields \cite{BF,iw} and for fermions in
AdS${}_4$ \cite{BF}). For any real mass and any $d \ge 2$, one may choose boundary
conditions that make our fermions stable at the level
of linear perturbations; formally, the condition $(m_\psi \pm
\frac{1}{2})^2 \ge 0$  is analogous to the Breitenlohner-Freedman
bound \cite{BF} for scalars $m^2_\phi \geq -(d-1)^2/4$. For $m^2_\psi
\geq 1/4$, only the faster falloff mode is normalizeable, so
boundary conditions must fix the coefficient $\alpha_\psi$ of the
slow falloff mode. For $0 \leq m^2_\psi < 1/4$, all modes are
normalizeable and more general boundary conditions are allowed. This
is directly analogous to the situation for scalars, where general
boundary conditions are permitted in the range $m^2_{BF} \leq
m^2_\phi <m^2_{BF}+1$.  However, for $d$ odd, the only
Lorentz-invariant derivative-free boundary conditions for a theory with a single
fermion are $\alpha_\psi =0$ or $\beta_\psi =0$.

For the examples of supersymmetry studied here, fermion and scalar
masses $(m_\psi,m_\phi)$ are related in all dimensions by
\begin{equation}
\label{corresp} m^2_{\phi, \pm}(m) = m^2_\psi \pm
\frac{m_\psi}{\ell} -\frac{d(d-2)}{4 \ell^2} \,,
\end{equation}
where we have restored factors of the AdS radius
$\ell$ and the $\pm$ denotes the fact that two scalar masses are
typically allowed for a given fermion mass. This formula also holds
in $d=2$ \cite{st}.  Our results for fermion boundary conditions at
mass $m_\psi$ typically agree with those for scalars at mass
$m_{\phi, \pm}$ when $m_\psi$ and $m_\phi$ satisfy (\ref{corresp}).
The one exception occurs for $m_\psi = \mp 1/(2\ell)$, which implies
$m^2_{\phi, \pm} = m^2_{BF}$ but $m^2_{\phi, \mp} = m^2_{BF}+1/\ell^2$.
Since the slow fall-off scalar mode is normalizeable for $m^2_{\phi,
\pm}$, but not for $m^2_{\phi, \mp}$,  it is clear that the fermion
cannot agree with both scalars.  In fact, the slow fall-off fermion
modes fail to be normalizeable in the standard inner
product\footnote{Though it might be interesting to reexamine this
issue using the techniques of \cite{cm}.}.  Thus, there are no supersymmetric multi-trace boundary conditions when the BF bound is saturated.
  This is consistent with {\it i\,}) the results of \cite{Amsel2006}, which found that for a single scalar at the BF bound, the Witten-Nester proof of the positive energy theorem does not apply unless one turns off the logarithmic mode and {\it ii\,}) the results of \cite{hlrs}, which argued (in the context of maximal gauged supergravity on AdS$_5$) that turning on the logarithmic branch leads to energies unbounded
  below.

We used such results to classify boundary conditions which preserve
supersymmetry (either a so-called Poincar\'e superalgebra involving
half of the supercharges or the full superalgebra) for certain
choices of field content.  In general, linear boundary conditions can
preserve only the Poincar\'e subalgebra of supercharges.  The same
is true of boundary conditions which would correspond to
deformations of a dual field theory involving an integer number of
traces.  Exceptions occur for special values of the fermion masses,
and for $d=3$ due to the nature of the AdS${}_3$ chiral
supermultiplet.  These exceptions are summarized in Table~\ref{tab2}.
For $d=4, m=0$, our results reduce to those of \cite{Marolf2006} and
yield the boundary conditions of \cite{BF} in a suitable limit.

\begin{table}
\begin{tabular}{|c|c|c|}
\hline
$d$ & \quad $m_\psi$ \quad &  \# Traces \\
\hline
\,3\, & \quad any value between $-\frac{1}{2}$  and $\frac{1}{2}$ \quad & 2  \\
\hline
\,4\, & 0 & 2 \\
 & $\pm 1/3$ & 3 \\
 \hline
\,5\, & 0 & 2 \\
\hline
\end{tabular}
\vspace{.5cm}
\caption{\label{tab2} Cases with integer number of traces and superconformal symmetry. Note that in $d=4$ such cases arise for both double-trace deformations ($m_\psi = 0$) and triple-trace deformations ($m_\psi = \pm 1/3$).}
\vspace{.2cm}
\end{table}

It may be interesting to perform a similar analysis including the
effects of backreaction, to investigate general boundary conditions
for vector and graviton supermultiplets, or to consider extended
supersymmetry.  However, of most interest would be a comparison with
a classification of supersymmetric deformations of a dual field
theory.  We close by discussing the details of  10-dimensional IIB supergravity
on AdS${}_5 \times S^5$, 11-dimensional supergravity on AdS${}_4 \times S^7$, and 10-dimensional IIA supergravity on AdS${}_4 \times
{\mathbb{CP}}^3$.  We then draw conclusions for the corresponding dual
theories; i.e., for ${\cal N} = 4$ super Yang-Mills in 3+1
dimensions \cite{M} and for the theories described in
\cite{BaggerLambert,ABJM}.

The case of AdS${}_5 \times S^5$ can be dealt with quickly.  From
\cite{AdS5S5}, we see that after Kaluza-Klein reduction on the
$S^5$, all spin 1/2 fields have AdS${}_5$ masses (in our notation) with magnitude
greater or equal to $1/2$. There are no allowed deformations of
boundary conditions for spin-1/2 fields, and thus no supersymmetric
deformations of boundary conditions of the type discussed here.
While we have not studied the spin-3/2 fields, for $d=5$ one does
not expect to be able to deform boundary conditions associated with
either vector or tensor fields in a Lorentz-invariant manner without
introducing ghosts \cite{iw,mr,cm}.  As a result, deformations of
the spin-3/2 boundary conditions are unlikely to be allowed, and
supersymmetric deformations will certainly be forbidden.  We
conclude that there are no  (relevant or marginal) multi-trace
deformations of the dual ${\cal N} = 4$ super Yang-Mills theory
which preserve even ${\cal N} = 1 $ Poincar\'e supersymmetry on the
boundary.

Let us now consider AdS${}_4 \times S^7$ and AdS${}_4 \times
\mathbb{CP}^3$. From \cite{AdS4S7} and \cite{AdS4CP3} we see that
these theories do contain fermions with masses $|m| < 1/2$.
Let us discuss the $S^7$ case for definiteness, though the $\mathbb
{CP}^3$ case is similar. For
AdS${}_4 \times S^7$, there is a single SO(8) multiplet of spin-1/2
fields in the desired mass range:  the 56${}_s$ representation of
SO(8) with mass $m=0$.  In addition, there are two SO(8) multiplets
(35${}_v$ and 35${}_c$) of conformally coupled scalars. Choosing an
${\cal N}=1$ super-Poincar\'e algebra on the boundary, we may
assemble from these fields 35 pairs of boundary superfields
$\Phi_\pm$ as described in section~\ref{d4}.  Allowed deformations
with integer numbers of traces are characterized by polynomials in
the 35 $\Phi_+$. Since each superfield has conformal dimension $1$, there are no relevant deformations and the marginal
deformations are labeled by the
$\binom{35}{2} = 595$ quadratic monomials formed from these fields.
I.e., there is a 595-dimensional manifold of conformal theories
connected by double-trace deformations. Since our superfields do not form a
well-defined SO(8) representation, none of these deformations will
preserve SO(8) symmetry (as is expected since we singled out an
${\cal N}=1$ subalgebra of the full supersymmetries).  Analogous reasoning leads to
a (somewhat smaller) manifold of conformal theories dual to
AdS${}_4\times \mathbb {CP}^3$.  We reached this conclusion ignoring
all bulk interactions, but from e.g. \cite{Amsel2006, hmtz} and the
fact that the theory has a symmetry that changes the sign of the
relevant scalars, one can see that including interactions will not
change this analysis.  The symmetry implies that the potential is even and forbids certain
logarithms that might otherwise be problematic for $m^2_\phi = -2$.

Strictly speaking, the above conclusions hold only at large $N$ and
one should ask if our marginal deformations might become relevant or
irrelevant at finite $N$.  While we cannot rule this out, from the
analysis of \cite{GM} it is clear that our deformations remain
marginal when leading $1/N$ corrections are included. Indeed, the
manifest AdS isometry appears to control bulk perturbation theory in
$G_N$ to all orders, so that our deformations remain exactly
marginal at all orders in the $1/N$ expansion.

\begin{acknowledgments}
This work was supported in part by the US National Science
Foundation under Grant No.~PHY05-55669, and by funds from the
University of California. D.M. thanks the Tata Institute for
Fundamental Research and the International Center for Theoretical
Sciences for their hospitality and support during the final stages
of this project.
\end{acknowledgments}

\appendix

\section{The Dirac equation in anti-de Sitter spacetime}
\label{revion2}

 In this appendix, we reduce the Dirac equation in
AdS${}_d$ to a set of coupled, first order differential equations
which may then be decoupled and easily solved.  The Dirac equation
in static, spherically symmetric spacetimes has been studied in
e.g., \cite{ion1, finster, ion2}, and in particular we now review
the results of \cite{ion2}.

The metric for AdS${}_d$ takes the form
\begin{equation}
\label{metricdh} ds^2 = -h(r) dt^2 +h^{-1}(r) dr^2 +r^2
d\Omega^2_{d-2} \,,
\end{equation}
where $h(r) = 1+r^2$.
Following \cite{ion2}, we define the ``Cartesian'' coordinates
\begin{eqnarray}
x^1 = r \cos \theta_1 \sin \theta_2 \ldots \sin \theta_{d-2}, & \ \
x^2 = r \sin \theta_1 \sin \theta_2 \ldots \sin \theta_{d-2},  \ \
 \dots,  \ \
x^{d-1} = r \cos{\theta_{d-2}} \ \cr {\rm with} & \ \ r^2 =
\sum^{d-1}_{k = 1} (x^k)^2 \equiv x^k x^k \,.
\end{eqnarray}
Then the spatial part of the metric (\ref{metricdh}) can be written
as
\begin{equation}
g_{ij} = \delta_{ij} - \frac{1}{r^2} \left(1 - \frac{1}{h} \right)
x^i x^j \,, \qquad i,j,k,\ldots = 1, 2, \ldots, d-1 \,,
\end{equation}
which leads us to choose the orthonormal frame
\begin{equation}
e^{\hat 0} = \sqrt{h} \, dt, \ \ e^{\hat k} =  dx^k - \frac{1}{r^2}
\left(1-\frac{1}{\sqrt{h}} \right) x^k x^j dx^j \,.
\end{equation}
As noted in \cite{finster}, for static, spherically symmetric
spacetimes the connection term $\gamma^a \Gamma_a$ that appears in the Dirac equation can be
computed from the simple formula
\begin{equation}
\label{simple} \gamma^a \Gamma_a = \frac{1}{2 \sqrt{-g}} \, \partial_a
(\sqrt{-g} \, \gamma^a) \,,
\end{equation}
where $g = \det g_{ab}$.
Using this result, the Dirac equation can be written as
\begin{equation}
\label{diracd} \frac{1}{\sqrt{h}} \gamma^{\hat 0} \partial_t \psi +
\frac{\sqrt{h}}{r^2} x^k \gamma^{\hat k}
\left[\left(1-\frac{1}{\sqrt{h}}\right) \left(x^j \partial_j
+\frac{d-2}{2}\ \right)+ \frac{r h'}{4 h} \right] \psi +\gamma^{\hat
k} \partial_k  \psi  -m \psi = 0 \,.
\end{equation}
The next step is to define the ``angular momentum'' operator $L_{ij}
= -i \left(x^i \partial_j - x^j \partial_i \right)$ and the Lorentz
generator $S^{i j } = \frac{1}{2} \gamma^{[\hat i} \gamma^{\hat j
]}$. Then, one can show that
\begin{equation}
\gamma^{\hat k} \partial_k = \frac{i}{r^2} x^k \gamma^{\hat k} S^{i
j} L_{ij} +\frac{1}{r^2} x^k \gamma^{\hat k} x^j \partial_j\,,
\end{equation}
which, upon rescaling $\psi = r^{-\frac{d-2}{2}} h^{-\frac{1}{4}}
\tilde \psi$ allows (\ref{diracd}) to be rewritten as
\begin{equation}
\label{diracd2} \gamma^{\hat 0} \partial_t \tilde \psi -
\frac{\sqrt{h}}{r^2} x^k \gamma^{\hat k} \left( \frac{d-2}{2}-i
S^{ij}L_{ij} \right) \tilde \psi +\frac{h}{r^2} x^ k \gamma^{\hat k}
x^j \partial_j \tilde \psi  -m \sqrt{h} \tilde \psi = 0 \,.
\end{equation}

Let us first suppose that $d =2 n $ is even.  Then we choose an
explicit gamma matrix representation
\begin{equation}
\label{gammaeven} \gamma^{\hat 0} =  \left(
\begin{matrix}
-i I_{2^{n-1}} & 0 \\
0 & i I_{2^{n-1}}
\end{matrix}
\right) , \quad \gamma^{\hat k} = \left(
\begin{matrix}
0 & -i\tau^{k}
\\ i\tau^{k} & 0
\end{matrix}\right)\,, \quad k = 1, 2, \ldots, 2n-1 \, ,
\end{equation}
where $\tau^k$ are $2^{n-1} \times 2^{n-1}$ matrices satisfying $
\{\tau^i, \tau^j\} = 2 \delta^{ij}$  and $I_n$ is the $n \times n$
identity matrix. In  four dimensions, the $\tau^k$ are just the
standard Pauli matrices
\begin{equation}
\label{pauli}
\sigma^1 =  \left(
\begin{matrix}
0 & 1 \\
1 & 0
\end{matrix}
\right), \quad \sigma^2 =  \left(
\begin{matrix}
0 & -i \\
i & 0
\end{matrix}
\right), \quad \sigma^3 =  \left(
\begin{matrix}
1 & 0 \\
0 & -1
\end{matrix}
\right) \, .
\end{equation}
In higher dimensions, the $\tau^k$ can be constructed
from tensor products of Pauli matrices (see e.g. \cite{guma}),
though we will not need the explicit expressions here.  To proceed further, we separate variables by making an
ansatz for solutions of the Dirac equation
\begin{equation}
\label{ansatzeven} \tilde \psi^\pm(t, r, \theta_i) = \left(
\begin{matrix}
i G^{\pm} (r) \mathscr{Y}^\pm_K (\theta_i)  \\
F^\pm (r)  \mathscr{Y}^\mp_K(\theta_i)
\end{matrix}
\right) \, e^{-i \omega t} \,,
\end{equation}
where $\mathscr{Y}^\pm_K$ are $2^{n-1}$-component spinor spherical
harmonics \cite{guma}.  These spinors satisfy
\begin{equation}
K  \mathscr{Y}^\pm_K  = \pm (l+n-1) \mathscr{Y}^\pm_K , \ \ {\rm
for} \ \ K \equiv \frac{d-2}{2}-\frac{i}{2} \tau^i \tau^j L_{ij}.
\end{equation}
Here $l = 0,1,2, \dots $ is the orbital angular momentum quantum
number.  In $d =4$, the operator $K$ takes the familiar form $(1+
\vec{\sigma} \cdot \vec{L})$, with $\vec{L} = -i \vec{r} \times
\vec{\partial}$ the usual angular momentum operator. In addition,
the spinor harmonics have the property $\frac{1}{r} x^j \tau^j \,
\mathscr{Y}^\pm_K =  \mathscr{Y}^\mp_K$, and the spinors $\psi^\pm$
are parity eigenstates, i.e. under parity $\psi^\pm \to \pm (-1)^l
\psi^\pm$. Substituting (\ref{ansatzeven}) into (\ref{diracd2}), we
obtain the coupled differential equations
\begin{equation}
h \frac{d F^\pm}{dr} = -\omega G^\pm \mp \frac{ k \sqrt{h}}{r} F^\pm
-m \sqrt{h} G^\pm, \ \ \  h \frac{d G^\pm}{dr} = \omega F^\pm \pm
\frac{ k \sqrt{h}}{r} G^\pm -m \sqrt{h} F^\pm \,,
\end{equation}
where $k = l +n-1$.

Now consider odd spacetime dimension $ d = 2n-1$.  We choose the
gamma matrix representation $\gamma^{\hat 0} = -i \tau^{2n - 1},
\, \gamma^{\hat k} = -i \tau^{2n-1} \tau^k, \, k = 1,\ldots,
2n-2$ where the $\tau^k$ are the same matrices referred to above for
the $d = 2n$ case. We make the ansatz
\begin{equation}
\label{ansatzodd1} \tilde \psi^+ = (i G^+ \mathscr{Y}^+_{K,1} +F^+
\mathscr{Y}^-_{K,1}), \ \ \tilde \psi^- = (i G^- \mathscr{Y}^-_{K,2}
+F^- \mathscr{Y}^+_{K,2})
\end{equation}
where $\mathscr{Y}^\pm_{K,p}$ ($p = 1,2$) are $2^{n-1}$-component spinor
spherical harmonics \cite{guma} satisfying
\begin{equation}
K  \mathscr{Y}^\pm_{K,p} =  \left( \frac{d-2}{2}-\frac{i}{2} \tau^i
\tau^j L_{ij} \right)  \mathscr{Y}^\pm_{K,p} = \pm
\left(l+\frac{2n-3}{2}\right)  \mathscr{Y}^\pm_{K,p} \,,
\end{equation}
\begin{equation}
\frac{1}{r} x^k \tau^k \mathscr{Y}^\pm_{K,p} = \mathscr{Y}^\mp_{K,p}
\,, \ \ {\rm and} \ \ \gamma^{\hat 0} \mathscr{Y}^\pm_{K,p} =  \pm i
(-1)^p \, \mathscr{Y}^\pm_{K,p} \,.
\end{equation}
Under parity, we again have $\psi^\pm \to \pm (-1)^l \psi^\pm$.
Using these relations in (\ref{diracd2}), we obtain the coupled
differential equations
\begin{equation}
h \frac{d F^\pm}{dr} = -\omega G^\pm \mp \frac{ k \sqrt{h}}{r} F^\pm
-m \sqrt{h} G^\pm , \ \  h \frac{d G^\pm}{dr} = \omega F^\pm \pm
\frac{ k \sqrt{h}}{r} G^\pm -m \sqrt{h} F^\pm \,,
\end{equation}
where $k = l +(2n-3)/2$.  We observe that these equations take
exactly the same form as in the even dimensional case.

To summarize, for any $d \geq 2$ we have reduced the Dirac equation
to a system of coupled ODEs
\begin{eqnarray}
- \frac{d F^\pm}{dr_*}  \mp k \csc r_* F^\pm -m \sec r_* G^\pm &=& \omega G^\pm  \\
 \frac{d G^\pm}{dr_*}  \mp k \csc r_*  G^\pm +m \sec r_* F^\pm  &=& \omega F^\pm\,,
\end{eqnarray}
where $k = l +(d-2)/2$ and we have defined a new radial coordinate
$r_*  = \tan^{-1} r$ whose range is $ [0, \pi/2)$.  In particular, the pair $F^+, G^+$ are coupled together, as are the pair
 $F^-, G^-$, but there is no mixing between the two pairs.
To solve these equations, we must first decouple them with a clever
trick, following \cite{ion1}. One first expresses the system in
matrix form
\begin{equation}
H_\pm \left(
\begin{matrix}
F^\pm \\
G^\pm
\end{matrix}
\right) = \omega \left(
\begin{matrix}
F^\pm \\
G^\pm
\end{matrix}
\right) \,, \ \ {\rm where} \ \ H_\pm = \left(
\begin{matrix}
m \sec r_* && \frac{d}{d r_*}\pm k \csc r_* \\
-\frac{d}{d r_*}\pm k \csc r_* && - m \sec r_*
\end{matrix}
\right)
\end{equation}
is the ``Hamiltonian.''  One then defines the unitary matrix $U =
\left(
\begin{matrix}
\cos \frac{r_*}{2} && \sin \frac{r_*}{2} \\
\sin \frac{r_*}{2} && -\cos \frac{r_*}{2}
\end{matrix}
\right)$ and performs the rotation $\left(
\begin{matrix}
\hat F^\pm \\
\hat G^\pm
\end{matrix}
\right) \equiv U \left(
\begin{matrix}
F^\pm \\
G^\pm
\end{matrix}
\right)\,$ . The rotated system then satisfies
\begin{equation}
\label{derotate} \hat H_\pm \left(
\begin{matrix}
\hat F^\pm \\
\hat G^\pm
\end{matrix}
\right) = \left(\omega-\frac{1}{2} \right) \left(
\begin{matrix}
\hat F^\pm \\
\hat G^\pm
\end{matrix}
\right) \,,
\end{equation}
where $\hat H_\pm = \left(
\begin{matrix}
m \mp k && -\frac{d}{d r_*}+ W_
\pm \\
\frac{d}{d r_*} + W_\pm && - (m \mp k)
\end{matrix}
\right) \,, \quad W_\pm = m \tan r_* \pm k \cot r_* \,$. Acting
again with $\hat H_\pm$, we have
\begin{equation}
\left(
\begin{matrix}
-\frac{d^2 \hat F^\pm}{d r^2_*} + \left(W^2_\pm -\frac{d W_\pm}{d r_*} +(m \mp k)^2 \right) \hat F^\pm \\
-\frac{d^2 \hat G^\pm}{d r^2_*}  + \left(W^2_\pm +\frac{d W_\pm}{d
r_*} +(m \mp k)^2 \right) \hat G^\pm
\end{matrix}
\right) = \left(\omega -\frac{1}{2} \right)^2 \left(
\begin{matrix}
\hat F^\pm \\
\hat G^\pm
\end{matrix}
\right) \,.
\end{equation}
Next, let us change radial coordinates again to $x = \pi/2 -r_*$.
Then $x$ has range $(0, \pi/2]$, with the conformal boundary
corresponding to $x = 0$.
Thus, we obtain the decoupled second order differential equations
\begin{eqnarray}
\label{deF}
-\frac{d^{\,2} \hat F^\pm}{d x^2}+  \left(\frac{\nu_-^2-1/4}{\sin^2 x} +\frac{\sigma_\pm^2 -1/4}{\cos^2 x} \right) \hat F^\pm &=& \tilde \omega^2  \hat F^\pm \\
\label{deG} -\frac{d^{\,2} \hat G^\pm}{d x^2}+
\left(\frac{\nu_+^2-1/4}{\sin^2 x} +\frac{\sigma_\mp^2 -1/4}{\cos^2
x} \right) \hat G^\pm &=& \tilde \omega^2  \hat G^\pm
\end{eqnarray}
where we have defined $\tilde \omega = \omega - 1/2$,
$\nu_\pm^2-\frac{1}{4} = m(m \pm 1)$, and $ \sigma_\pm^2-\frac{1}{4}
= k(k \pm 1)$. Solutions to these differential equations are
discussed in appendix~\ref{normpsi}.


\section{Normalizeable modes for Dirac fermions}
\label{normpsi}

We now analyze normalizeability for massive Dirac fermions using the
standard inner product
\begin{equation}
\label{structure}
\sigma_\Sigma(\delta_1 \psi, \delta_2 \psi) =
i \int d^{d-1}x \sqrt{g_\Sigma} \, t_a \left(\overline{\delta_1
\psi} \gamma^a \delta_2 \psi - \overline{\delta_2 \psi} \gamma^a
\delta_1 \psi \right)
\end{equation}
between linearized solutions.  Here $\Sigma$ is a hypersurface
defined by $ t = constant$ with unit normal $t^a$ and $g_\Sigma$ is
the determinant of the induced metric on $\Sigma$.  We will rely
heavily on the treatment of
the AdS${}_d$ Dirac equation in \cite{ion2} (see the summary in appendix \ref{revion2}), and in particular on
equations \eqref{deF}, \eqref{deG}.

Assuming that $\nu_\pm, \sigma_\pm \geq 0$ (and for now $m \geq 0$),
we have $\nu_- = \left|m-\frac{1}{2}\right|\,, \, \nu_+ = m+\frac{1}{2}$,
while $\sigma_- = \left|l+\frac{d-3}{2}\right|\,,\, \sigma_+ = l +
\frac{d-1}{2}$. Inserting the ansatz (\ref{ansatzeven}) or
(\ref{ansatzodd1}) into (\ref{structure}) and assuming that the
spinor spherical harmonics are properly normalized yields
 \begin{equation}
 \label{usualIP}
 \sigma_\Sigma(\delta_1 \psi^\pm, \delta_2 \psi^\pm)
= -i \int^{\pi/2}_0 dx \left( (\delta_1 G^\pm)^* \delta_2 G^\pm +
(\delta_1 F^\pm)^* \delta_2 F^\pm \right) - \left(\delta_1
\leftrightarrow \delta_2 \right) \,.
 \end{equation}
We see that requiring the inner product to be finite is the
same as requiring $\delta F^\pm, \delta G^\pm$ to be square
integrable on $L^2(x \in [0, \pi/2])$. Since the rotation by $U$ to
the new radial functions $\hat F^\pm, \hat G^\pm$ is unitary, this
is further equivalent to square integrability of $\delta \hat F^\pm,
\delta \hat G^\pm$ .

It is useful to observe that the equations of motion for fermions
(\ref{deF}),(\ref{deG}) have been put in the same general form as
that used for scalar, vector, and tensor fields in \cite{iw}. Hence,
we can take advantage of the analysis already performed in that
reference. For even $d$, the $\sigma_\pm$ are non-integer, and the
general solutions to (\ref{deF}), (\ref{deG}) are hypergeometric
functions
\begin{eqnarray}
\hat F^\pm  = &B^\pm_1& (\sin x)^{\nu_- +1/2} (\cos x)^{\sigma_\pm
+1/2} \, {}_2F_1 (\zeta^{\tilde \omega}_{\nu_- , \sigma_\pm},
\zeta^{-\tilde \omega}_{\nu_- , \sigma_\pm},1+\sigma_\pm, \cos^2 x)
\nonumber \\ &+& B^\pm_2 (\sin x)^{\nu_- +1/2} (\cos x)^{-\sigma_\pm
+1/2} \, {}_2F_1 (\zeta^{\tilde \omega}_{\nu_- , -\sigma_\pm},
 \zeta^{-\tilde \omega}_{\nu_- , -\sigma_\pm},1-\sigma_\pm, \cos^2 x)
 \end{eqnarray}
 \begin{eqnarray}
\hat G^\pm  = &C^\pm_1& (\sin x)^{\nu_+ +1/2} (\cos x)^{\sigma_\mp
+1/2} \, {}_2F_1 (\zeta^{\tilde \omega}_{\nu_+ , \sigma_\mp},
\zeta^{-\tilde \omega}_{\nu_+ , \sigma_\mp}, 1+\sigma_\mp, \cos^2 x)
\nonumber \\ &+& C^\pm_2 (\sin x)^{\nu_+ +1/2} (\cos x)^{-\sigma_\mp
+1/2} \, {}_2F_1 (\zeta^{\tilde \omega}_{\nu_+ , -\sigma_\mp},
 \zeta^{-\tilde \omega}_{\nu_+ , -\sigma_\mp},1-\sigma_\mp, \cos^2 x)
 \end{eqnarray}
where $\zeta^\omega_{\nu, \sigma} = \frac{\nu+\sigma+1+\omega}{2}$.
Near the origin, $x \sim \pi/2$, we have $\hat F^\pm  = B^\pm_2
(\cos x)^{-\sigma_\pm +1/2}+ \ldots,$ $\hat G^\pm  = C^\pm_2 (\cos
x)^{-\sigma_\mp +1/2}+ \ldots$. Thus $\hat F^\pm, \hat G^\pm$ are
not square integrable near the origin if $\sigma_\pm \geq 1$. This
inequality is always satisfied for even $d$, except for the
cases $d=2$ ($\sigma_\pm = 1/2$) and $d = 4, l= 0 $ ($\sigma_- = 1/2$).

For odd $d$,  the $\sigma_\pm$ are integers, and the second linearly
independent solution to $\hat F^\pm, \hat G^\pm$ is modified (see
\cite{iw}).  For $\sigma_\pm \neq 0$, the solutions behave near the
origin as $\hat F^\pm  \propto  B^\pm_2 (\cos x)^{-\sigma_\pm
+1/2}\left( (\cos x)^{-2\sigma_\pm} + \ldots \right)$, $\hat G^\pm
\propto  C^\pm_2 (\cos x)^{-\sigma_\mp +1/2}\left( (\cos
x)^{-2\sigma_\mp} + \ldots \right)$. Here $\sigma_\pm \geq 1$, and
so $\hat F^\pm, \hat G^\pm$ are not square integrable near the
origin.  When $d = 3, l = 0$ we have $\sigma_- = 0$ and near the
origin $\hat F^-  = B^\pm_2 (\cos x)^{1/2}\log (\cos^2 x) + \ldots$,
$\hat G^+  = C^\pm_2 (\cos x)^{1/2}\log (\cos^2 x) + \ldots$. These
solutions are square integrable near $x\sim\pi/2$.

We have seen that square integrability requires $B^\pm_2 = 0 =
C^\pm_2$, except in the cases $\sigma_\pm = 0, 1/2$.  However, as
explained in \cite{iw}, the solutions in these special cases are
actually not acceptable, as they correspond to solutions of an
equation with a $\delta$-function source.  This is a result of
having removed the origin when we chose spherical coordinates.  So,
in all cases we set $B^\pm_2 = 0 = C^\pm_2$.  We also note here that
the constants $B^+_1 , C^+_1$  ($B^-_1 , C^-_1$) are not independent
because the functions $\hat F^+, \hat G^+$ ($\hat F^-, \hat G^-$)
are coupled through the first order differential equation
(\ref{derotate}) \cite{ion1}.  In fact, one can obtain the
consistency conditions
\begin{equation}
\frac{C^+_1}{B^+_1} = -\frac{2(2l+d-1)}{2l+d-2-2m-2 \tilde \omega}
\,, \qquad \frac{B^-_1}{C^-_1} = \frac{2(2l+d-1)}{2l+d-2+2m-2\tilde
\omega} \,.
\end{equation}

Now consider the behavior near infinity, $x \to 0$.  For this, it is
best to write the hypergeometric functions as functions of $\sin^2
x$.  For example, when $\nu_\pm \neq 0,1,2, \ldots$, we have
\begin{eqnarray}
\hat F ^\pm = &B^\pm_1& (\cos x)^{\sigma_\pm +1/2} (\sin x)^{-\nu_-
+1/2}\Bigg[\frac{\Gamma(1+\sigma_\pm)
\Gamma(\nu_-)}{\Gamma(\zeta^{\tilde \omega}_{\nu_- , \sigma_\pm})
 \Gamma(\zeta^{-\tilde \omega}_{\nu_- , \sigma_\pm})}
{}_2F_1(\zeta^{\tilde \omega}_{-\nu_- , \sigma_\pm}, \zeta^{-\tilde \omega}_{-\nu_- , \sigma_\pm},1-\nu_-, \sin^2 x) \nonumber\\
&+&\frac{\Gamma(1+\sigma_\pm) \Gamma(-\nu_-)}{\Gamma(\zeta^{\tilde
\omega}_{-\nu_- , \sigma_\pm})
 \Gamma(\zeta^{-\tilde \omega}_{-\nu_- , \sigma_\pm})} (\sin x)^{2 \nu_-}
 {}_2F_1(\zeta^{\tilde \omega}_{\nu_- , \sigma_\pm}, \zeta^{-\tilde \omega}_{\nu_- , \sigma_\pm},1+\nu_-, \sin^2 x) \Bigg] \,,
 \end{eqnarray}
with the corresponding expression for $\hat G^\pm$ given by
exchanging $\nu_- \to \nu_+\,, \sigma_\pm \to \sigma_\mp$.  The
transformations of the hypergeometric functions for the remaining
cases of integer $\nu_\pm$ are given in \cite{iw}.  Near the
boundary, the leading terms in all cases are
\begin{eqnarray}
\hat F^\pm  &\sim& B^\pm_1 \frac{\Gamma(1+\sigma_\pm)
\Gamma(\nu_-)}{\Gamma(\zeta^{\tilde \omega}_{\nu_- , \sigma_\pm})
 \Gamma(\zeta^{-\tilde \omega}_{\nu_- , \sigma_\pm})} (\sin x)^{-\nu_-+1/2}+ \ldots\\
\hat G^\pm &\sim& C^\pm_1 \frac{\Gamma(1+\sigma_\mp)
\Gamma(\nu_+)}{\Gamma(\zeta^{\tilde \omega}_{\nu_+ , \sigma_\mp})
 \Gamma(\zeta^{-\tilde \omega}_{\nu_+ , \sigma_\mp})} (\sin x)^{-\nu_+ + 1/2}+ \ldots
\end{eqnarray}
We examine the mass ranges with distinct behavior in turn.

$\mathbf{ m\geq 3/2:}$ This corresponds to $\nu_- \geq 1, \nu_+ \geq
2$. Then $\hat F^+$ is not square integrable unless
$\Gamma(\zeta^{\tilde \omega}_{\nu_- , \sigma_+})$ or
$\Gamma(\zeta^{-\tilde \omega}_{\nu_- , \sigma_+})$ diverges, i.e.
\begin{equation}
\tilde \omega = \mp (2 n +1+\nu_- + \sigma_+) \,, \quad n = 0,1,2,
\ldots \,.
\end{equation}
Since $\hat F^+, \hat G^+$ were coupled in the original Dirac
equation, this also fixes $\tilde \omega$ in the $\hat G^+$
solution.  Then $\Gamma(\zeta^{\pm \tilde \omega}_{\nu_+ ,
\sigma_-}) = \Gamma(-n)$ diverges, and so this ensures that $\hat
G^+$ is also square integrable. Similarly, $\hat F^-$ is not square
integrable unless $\Gamma(\zeta^{\tilde \omega'}_{\nu_- ,
\sigma_-})$ or $\Gamma(\zeta^{-\tilde \omega'}_{\nu_- , \sigma_-})$
diverges, i.e.
\begin{equation}
\tilde \omega' = \mp (2 n' +1+\nu_- + \sigma_-) \,, \quad n' = 1,2,
\ldots \,.
\end{equation}
(Here we have been careful to note that one could choose independent
frequencies $\omega$ and $\omega'$ for the $\psi^+$ and $\psi^-$
solutions.) Since $\hat F^-, \hat G^-$ were coupled in the original
Dirac equation, this also fixes $\tilde \omega'$ in the $\hat G^-$
solution.  Then $\Gamma(\zeta^{\pm \tilde \omega'}_{\nu_+ ,
\sigma_+}) = \Gamma(-n'+1)$ diverges, and so this ensures that $\hat
G^-$ is also square integrable.

$\mathbf{  1/2 \leq m < 3/2:}$ This corresponds to $0 \leq \nu_- <
1, 1 \leq \nu_+ < 2$.  Then, $\hat F^+$ is square integrable for
all $\tilde \omega$, but we still must fix
\begin{equation}
\tilde \omega = \mp (2 n +1+\nu_+ + \sigma_-) \,, \quad n = 0,1,2,
\ldots
\end{equation}
to ensure that $\hat G^+$ is square integrable.  Similarly, $\hat
F^-$ is square integrable for all $\tilde \omega'$, but we still
must fix
\begin{equation}
\tilde \omega' = \mp (2 n' +1+\nu_+ + \sigma_+) \,, \quad n' =
0,1,2, \ldots
\end{equation}
to ensure that $\hat G^-$ is square integrable.

$\mathbf{ 0 \leq m < 1/2:}$ This corresponds to $0 < \nu_- \leq 1/2,
1/2 \leq \nu_+ < 1$.  In this case, $\hat F^\pm, \hat G^\pm$ are
square integrable for all $\tilde \omega$.

We have thus found that for $m\geq 1/2$, requiring the inner product
to be finite imposes a unique boundary condition.  In
terms of the original spinor fields $\psi$, we note (using
(\ref{ansatzeven})) that near infinity
\begin{equation}
\label{expand} \psi^\pm \sim \left(
\begin{matrix}
i \mathscr{Y}^\pm \\
\mathscr{Y}^\mp
\end{matrix}
\right) e^{-i \omega t} (\sin x)^{\frac{d-1}{2}} \hat F^\pm + \left(
\begin{matrix}
-i \mathscr{Y}^\pm \\
\mathscr{Y}^\mp
\end{matrix}
\right) e^{-i \omega t} (\sin x)^{\frac{d-1}{2}} \hat G^\pm \,,
\quad d \, \, \mathrm{even}
\end{equation}
and similarly for $d$ odd.  Expanding $\hat F^\pm, \hat G^\pm$ for
$x \to 0$ and using the frequency quantization conditions above, we
find that asymptotically
\begin{equation}
\psi \sim \beta (\sin x)^{\frac{d-1}{2}+m} + O\left((\sin
x)^{\frac{d+1}{2}+m}\right) \,,
\end{equation}
where the coefficient $\beta$ is a spinor depending on time and
angles on the $S^{d-2}$, but not on $x$.

For $0 \leq m < 1/2$, there will be a choice of boundary conditions
at infinity. Note that, unlike the scalar case, this mass range does
not depend on $d$. Using (\ref{expand}) and expanding $\hat F^\pm,
\hat G^\pm$ for $x \to 0$, we find that near infinity
\begin{equation}
\label{expand2} \psi \sim \alpha (\sin x)^{\frac{d-1}{2}-m} + \beta
(\sin x)^{\frac{d-1}{2}+m} + O\left((\sin
x)^{\frac{d+1}{2}-m}\right) \,,
\end{equation}
where the coefficients $\alpha, \beta$ are spinors depending only on
time and angles on the $S^{d-2}$.  Using the properties of the
spinor spherical harmonics under the action of the radial gamma
matrix $x^k \gamma^{\hat k }$ (see appendix \ref{revion2}), one can
verify that $P_- \alpha = 0$, $P_+ \beta = 0$ where we have defined
the radial gamma matrix projectors $P_\pm = \frac{1}{2} \left(1 \pm
\frac{1}{r} x^k \gamma^{\hat k } \right)$.

One may also work out the sub-leading terms which will be needed for
the study of supersymmetry in the main text.  The important step is
to rewrite the Dirac equation in terms of the unphysical metric
$\tilde g_{ab}$:
\begin{equation}
\gamma^a \nabla_a \psi-m\psi = \Omega \tgam^a \tnabla_a \psi
-\frac{d-1}{2} \, \tn_a \tgam^a \psi - m\psi = 0 \,.
\end{equation}
If we now insert the expansion (\ref{solexpand}) into the above
equation and collect terms we find
\begin{eqnarray}
\label{diracexpand} 0 = &-&m \left(1+ \tn_a \tgam^a\right) \alpha
\Omega^{\frac{d-1}{2}-m}- m \left(1-\tn_a \tgam^a\right) \beta
\Omega^{\frac{d-1}{2}+m} \nonumber
\\&+&\left[\tgam^a \tnabla_a \alpha - \left(m+(m-1) \tn_a \tgam^a \right)\alpha' \right] \Omega^{\frac{d+1}{2}-m}
\nonumber \\ &+&\left[\tgam^a \tnabla_a \beta - \left(m-(m+1) \tn_a
\tgam^a \right)\beta' \right] \Omega^{\frac{d+1}{2}+m}
+O(\Omega^{\frac{d+3}{2}-|m|}) \,.
\end{eqnarray}
Setting each term to zero leads to the relations
\eqref{projecta},\eqref{projectb} stated in section \ref{BCs}.

We conclude this appendix with a discussion of the relation to
\cite{iw}.  As noted in section \ref{BCs}, an alternative method of
analyzing allowed boundary conditions is to consider self-adjoint
extensions of an appropriate spatial wave operator.  In \cite{iw},
this analysis is performed for massive scalars and massless vectors
and tensors and it is shown that all such cases reduce to studying
the operator
\begin{equation}
\label{op} A = -\frac{d^2}{dx^2} +\frac{\nu^2-1/4}{\sin^2
x}+\frac{\sigma^2 -1/4}{\cos^2 x}
\end{equation}
on the Hilbert space $L^2([0, \pi/2], dx)$.  The results are
determined by $\nu^2$.  If $\nu^2 < 0$, then $A$ is unbounded below
and so does not admit a positive extension; this is the case for
scalars with $m^2 < m^2_{BF}$. If $\nu^2 \geq 0$,  $A$ is a positive
operator (and therefore there exists at least one positive
self-adjoint extension).   For $\nu^2 \geq 1$  there is a unique
self-adjoint extension that is automatically positive and so a unique linear boundary
condition at infinity. However, for
 $0 \leq \nu^2 < 1$ (e.g. $m^2_{BF} \leq m^2_\phi < m^2_{BF}+1$ for scalars) there is a family of such extensions
 corresponding to a choice of boundary conditions.
Wald and Ishibashi proceed to determine all possible linear boundary
conditions corresponding to positive self-adjoint extensions.

Since the same wave operator \eqref{op} appears in
\eqref{deF},\eqref{deG} and the inner product \eqref{usualIP} is the
same as above, we can apply the analysis of \cite{iw} directly to
any massive Dirac fermion:  The wave operators are symmetric on the
domain of smooth functions of compact support away from the origin,
$C_0^\infty (0, \pi/2)$, and are positive for $\nu_\pm^2 \geq 0$,
which implies that they admit at least one positive self-adjoint
extension. In terms of the mass $m$, this condition is equivalent to
\begin{equation}
\label{bffermion} \left(m\pm\frac{1}{2}\right)^2 \geq 0
\Longrightarrow m^2 \geq 0 \,.
\end{equation}
For $\nu^2_\pm <0$, the operators are unbounded below and therefore
do not admit a positive self-adjoint extension. The inequality
(\ref{bffermion}) is the analogue of the Breitenlohner-Freedman
bound for stability, though of course it is trivially satisfied for
real $m$. The case $m^2 = 1/4$ is analogous to saturating the BF
bound. For $m\geq 1/2$, the wave operators have a unique, positive
self-adjoint extension and so there is a unique linear boundary
condition at infinity. For $0 \leq m < 1/2$, there is a family of
self-adjoint extensions, and a choice of boundary conditions at
infinity. One could also follow the von Neumann prescription as in
\cite{iw} to classify positive self-adjoint extensions and the
corresponding boundary conditions.


\begin{thebibliography}{11}

\bibitem{M}
J.~M.~Maldacena,
 ``The large N limit of superconformal field
theories and supergravity,'' Adv.\ Theor.\ Math.\ Phys.\  {\bf 2}
(1998) 231, [arXiv:hep-th/9711200].

\bibitem{gkp}
S.~S.~Gubser, I.~R.~Klebanov, and A.~M.~Polyakov,
``Gauge theory
correlators from non-critical string theory,'' Phys.\ Lett.\ B {\bf
428}, 105 (1998) [arXiv:hep-th/9802109 ].

\bibitem{witten}
E.~Witten,
``Anti-de Sitter space and holography,'' Adv.\ Theor.\
Math.\ Phys.\ {\bf 2}, 253 (1998) [arXiv:hep-th/9802150].

\bibitem{kw}
I.~R.~Klebanov and E.~Witten,
``AdS/CFT correspondence and symmetry breaking,''
Nucl.\ Phys.\ B {\bf 556} (1999) 89-114, [arXiv:hep-th/9905104].

\bibitem{FG}
C.~Fefferman and C.~R.~Graham,
"Conformal Invariants", in {\it Elie Cartan et les Math$\acute{e}$matiques d'aujourd'hui, Ast$\acute{e}$risque}, 1985, 95.

\bibitem{witten1}
E.~Witten,
``Multi-trace operators, boundary conditions, and AdS/CFT
correspondence,'' [arXiv:hep-th/0112258].

\bibitem{BF}
P.~Breitenlohner and D.~Z.~Freedman, ``Stability in gauged extended
supergravity,'' Annals Phys.\  {\bf 144} (1982) 249; ``Positive
energy in anti-de Sitter backgrounds and gauged extended
supergravity,'' Phys.\ Lett.\ B {\bf 115} (1982) 197.

\bibitem{iw}
 A.~Ishibashi and R.~M.~Wald,
  ``Dynamics in non-globally hyperbolic static spacetimes III: anti-de Sitter
  spacetime,''
  Class.\ Quant.\ Grav.\  {\bf 21}, 2981 (2004)
  [arXiv:hep-th/0402184].

\bibitem{mr}
D.~Marolf and S.~Ross, ``Boundary conditions and dualities: vector
fields in AdS/CFT,'' JHEP {\bf 0611} (2006) 085,
[arxiv:hep-th/0606113].

\bibitem{cm}
G. Comp$\grave{\mathrm{e}}$re and D.~Marolf, ``Setting the boundary
free in AdS/CFT,'' arXiv:0805.1902 [hep-th].


\bibitem{Marolf2006}
  S.~Hollands and D.~Marolf,
  ``Asymptotic generators of fermionic charges and boundary conditions
  preserving supersymmetry,''
  Class.\ Quant.\ Grav.\ {\bf 24} 2301-2332, (2007)
  [arXiv:gr-qc/0611044].


\bibitem{hawking}
S.~W.~Hawking, ``The boundary conditions for gauged supergravity,''
Phys.\ Lett.\ B {\bf 126} 175, (1983).

\bibitem{in}
Y.~Igarashi and T.~Nonoyama, ``Supersymmetry and reflective boundary
conditions in anti-de Sitter spaces'' Phys.\ Rev.\ D {\bf 34} 1928
(1986).

\bibitem{st}
N.~Sakai and Y.Tanii, ``Supersymmetry in two-dimensional anti-de
Sitter space,'' Nucl.\ Phys.\ B {\bf 258}, 661 (1985).

\bibitem{bss}
  M.~Berkooz, A.~Sever, and A.~Shomer,
 ``Double-trace deformations, boundary conditions and spacetime
 singularities,''
  JHEP {\bf 0205} (2002) 034,
  [arXiv:hep-th/0112264].

\bibitem{ss}
   A.~Sever and A.~Shomer,
  ``A note on multi-trace deformations and AdS/CFT,''
  JHEP {\bf 0207} (2002) 027,
 [arXiv:hep-th/0203168].

 \bibitem{vN}
 D.~V.~Belyaev and P.~van Nieuwenhuizen,
  ``Rigid supersymmetry with boundaries,''
  JHEP {\bf 0804}, 008 (2008),
  arXiv:0801.2377 [hep-th];
   P.~van Nieuwenhuizen and D.~V.~Vassilevich,
  ``Consistent boundary conditions for supergravity,''
  Class.\ Quant.\ Grav.\  {\bf 22}, 5029 (2005)
  [arXiv:hep-th/0507172].


  \bibitem{hs}
  M.~Henningson and K.~Sfestos,
  ``Spinors and the AdS/CFT correspondence,''
  Phys.\ Lett.\ B {\bf 431} (1998) 63, [arXiv:hep-th/9803251].
  
  \bibitem{frolov}
  G.~E.~Arutyunov and S.~A.~Frolov,
  ``On the origin of supergravity boundary terms in the AdS/CFT correspondence,''
  Nucl.\ Phys.\ B {\bf 544} (1999) 576-589, [arXiv:hep-th/9806216].

  \bibitem{henneaux99}
  M.~Henneaux,
  ``Boundary terms in the AdS/CFT correspondence for spinor fields,''
  [arXiv:hep-th/9902137].

  \bibitem{hmtz}
M.~Henneaux, C.~Martinez, R.~Troncoso, and J.~Zanelli,
``Asymptotic behavior and Hamiltonian analysis of anti-de Sitter gravity coupled
to scalar fields,''
Annals Phys. {\bf 322} (2007) 824-848, [arXiv:hep-th/0603185].

\bibitem{Amsel2006}
A.~J.~Amsel and D.~Marolf,
``Energy bounds in designer gravity''
Phys.\ Rev.\ D {\bf 74}, 064006 (2006), [arXiv:hep-th/0605101].

\bibitem{thgh}
T.~Hertog and G.~T.~Horowitz,
``Designer gravity and field theory effective potentials,''
Phys.\ Rev.\ Lett.\ {\bf 94} (2005) 221301,
[arXiv:hep-th/0412169].

\bibitem{ion2}
    I.~I.~Cot$\breve{\rm a}$escu,
    ``Discrete quantum modes of the Dirac field in $AdS_{d+1}$ backgrounds,''
    Int.\ J.\ Mod.\ Phys.\ A {\bf 19} (2004) 2217-2232 [arXiv:gr-qc/0306127].

\bibitem{BaggerLambert}
 J.~Bagger and N.~Lambert,
  ``Three-algebras and N=6 Chern-Simons gauge theories,''
  arXiv:0807.0163 [hep-th].

\bibitem{ABJM}
O.~Aharony, O.~Bergman, D.~L.~Jafferis, and J.~Maldacena,
  ``N=6 superconformal Chern-Simons-matter theories, M2-branes and their
  gravity duals,''
  arXiv:0806.1218 [hep-th].

\bibitem{mr2}
D.~Marolf and S.~Ross, ``Reversing renormalization-group flows with AdS/CFT,'' JHEP {\bf 0805} (2008) 055, arXiv:0705.4642 [hep-th].

 \bibitem{is}
 E.~A.~Ivanov and E.~A.~Sorin,
 ``Superfield formulation of  OSp(1,4) supersymmetry,''
 J.\ Phys.\ A {\bf 13}, 1159 (1980).


\bibitem{dewit}
B.~de Wit and I.~Herger, ``Anti-de Sitter supersymmetry,''
[arXiv:hep-th/9908005].

\bibitem{Henneaux85}
M. Henneaux and C. Teitelboim, ``Asymptotically anti-de Sitter
spaces," Commun. Math. Phys. {\bf 98} (1985) 391.

\bibitem{pvn}
P.~van Nieuwenhuizen, ``An introduction to simple supergravity and
the Kaluza-Klein program''
 in {\it Les Houches 1983 Proceedings, Relativity, Groups, and Topology II} 823-932.

\bibitem{shuster}
E.~Shuster, ``Killing spinors and supersymmetry in AdS,'' Nucl.\
Phys.\ B {\bf 554} (1999) 198-214 [arXiv:hep-th/9902129].


\bibitem{at}
A.~Achucarro and P.~K.~Townsend,
``A Chern-Simons action for three-dimensional AdS supergravity theories,''
Phys.\ Lett.\ B {\bf 180} (1986) 89.

\bibitem{it}
    J.~M.~Izquierdo and P.~K.~Townsend,
    ``Supersymmetric spacetimes in $(2+1)$ AdS-supergravity models,''
    Class.\ Quant.\ Grav.\ {\bf 12} (1995) 895, [arXiv:gr-qc/9501018].

\bibitem{dkss}
N.~S.~Deger, A.~Kaya, E.~Sezgin, and P.~Sundell,
``Matter coupled $AdS_3$ supergravities and their black strings,''
Nucl.\ Phys.\  B {\bf 573} (2000) 275-290, [arXiv:hep-th/9908089].

\bibitem{dksst}
N.~S.~Deger, A.~Kaya, E.~Sezgin, P.~Sundell, and Y.~Tanii,
``$(2,0)$ Chern-Simons supergravity plus matter near the boundary of
$AdS_3$,''
Nucl.\ Phys.\ B {\bf 604} (2001) 343-366, [arXiv:hep-th/0012139].

\bibitem{hlrs}
 V.~E.~Hubeny, X.~Liu, M.~Rangamani, and S.~Shenker,
 ``Comments on cosmic censorship in AdS/CFT,''
JHEP {\bf 0412} (2004) 067, [arXiv:hep-th/0403198].


\bibitem{AdS5S5}
H.~J.~Kim, L.~J.~Romans, and P.~van Nieuwenhuizen,
``The mass spectrum of chiral N=2 D=10 supergravity on $S^5$,''
Phys.\ Rev.\ D {\bf 32} (1985) 389.

\bibitem{AdS4S7}
M.~Gunaydin and N.~P.~Warner,
``Unitary supermultiplets of Osp(8/4, ${\mathbb R}$) and the spectrum
of the $S^7$ compactification of eleven-dimensional supergravity,''
Nucl.\ Phys.\ B {\bf 272} (1986) 99.  

\bibitem{AdS4CP3}
B.~E.~W.~Nilsson and C.~N.~Pope,
``Hopf fibration of eleven-dimensional supergravity,''
Class.\ Quant.\ Grav.\ {\bf 1}, 499 (1984).

\bibitem{GM}
S.~S.~Gubser and I.~Mitra,
  ``Double-trace operators and one-loop vacuum energy in AdS/CFT,''
  Phys.\ Rev.\  D {\bf 67}, 064018 (2003)
  [arXiv:hep-th/0210093].

\bibitem{ion1}
    I.~I.~ Cot$\breve{\rm a}$escu,
    ``The Dirac particle on central backgrounds and the anti-de Sitter oscillator,''
    Mod.\ Phys.\ Lett.\ A {\bf 13} (1998) 2923-2936 [arXiv:gr-qc/9803042].


\bibitem{finster}
    F.~Finster, J.~Smoller, and S.~T.~Yau,
    ``Particle-like solutions of the Einstein-Dirac equations,''
    Phys.\ Rev.\ D {\bf 59} (1999) 104020 [gr-qc/9801079].

\bibitem{guma}
    X.~Y.~Gu and Z.~Q.~Ma,
    ``Exact solutions to the Dirac equation for a Coulomb potential in $D+1$ dimensions,''
    [arXiv:physics/02090391].


\end{thebibliography}
\end{document}